\documentclass[aps,prb,twocolumn,floats]{revtex4}
\usepackage{color,ulem,verbatim,float,wrapfig}
\usepackage{amssymb,amsbsy,amsmath,mathrsfs}
\usepackage{graphicx}
\usepackage{times}
\usepackage{color}
\usepackage{subfigure}
\usepackage{braket}
\usepackage{bbold}
\usepackage{commath}
\usepackage{setspace}
\usepackage{bm} 
\newcommand\bea{\begin{eqnarray}}
\newcommand\eea{\end{eqnarray}}
\newcommand\beq{\begin{equation}}
\newcommand\eeq{\end{equation}}
\newcommand\bib{\bibitem}

\newcommand{\non}{\nonumber}
\newcommand{\al}{\alpha}
\newcommand{\de}{\delta}

\newcommand{\ga}{\gamma}
\newcommand{\ep}{\epsilon}

\newcommand{\lm}{\lambda}
\newcommand{\si}{\sigma}
\newcommand{\ta}{\theta}
\newcommand{\om}{\omega}
\newcommand{\dg}{\dagger}
\newcommand{\ua}{\uparrow}
\newcommand{\da}{\downarrow}

\newcommand{\vn}{{\vec n}}
\newcommand{\vu}{{\vec u}}
\newcommand{\vv}{{\vec v}}
\newcommand{\vw}{{\vec w}}
\newcommand{\vk}{{\vec k}}
\newcommand{\vq}{{\vec q}}

\usepackage{soul}

\begin{document}

\title{Driven Hubbard model on a triangular lattice: tunable Heisenberg antiferromagnet 
with chiral three-spin term}
\author{Samudra Sur$^1$, Adithi Udupa$^1$ and Diptiman Sen$^{1,2}$}
\affiliation{$^1$Center for High Energy Physics, Indian Institute of Science, Bengaluru 560012, India \\
$^2$Department of Physics, Indian Institute of Science, Bengaluru 560012,
India}

\begin{abstract}
We study the effects of a periodically varying electric field on the Hubbard
model at half-filling on a triangular lattice. The electric field is
incorporated through the phase of the nearest-neighbor hopping amplitude via
the Peierls prescription. When the on-site interaction $U$ is much larger than
the hopping, the effective Hamiltonian $H_{eff}$ describing the spin sector
can be found using a Floquet perturbation theory. To third order in the
hopping, $H_{eff}$ is found to have the form of a Heisenberg antiferromagnet
with three different nearest-neighbor couplings $(J_\al,J_\beta,J_\ga)$ on
bonds lying along the different directions. Remarkably, when the periodic
driving does not have time-reversal symmetry, $H_{eff}$ can also
have a chiral three-spin interaction in each triangle, with the coefficient $C$
of the interaction having opposite signs on up- and down-pointing triangles.
Thus periodic driving which breaks time-reversal symmetry can simulate the 
effect of a perpendicular magnetic flux which is known to generate such a 
chiral term in the spin sector, even though our model does not have a
magnetic flux. The four parameters $(J_\al,J_\beta,J_\ga,C)$ depend on the
amplitude, frequency and direction of the oscillating electric field. We
then study the spin model as a function of these parameters using exact 
diagonalization and find a rich phase diagram of the ground state 
with seven different phases consisting of two kinds of ordered
phases (collinear and coplanar) and disordered phases. Thus periodic driving of
the Hubbard model on the triangular lattice can lead to an effective spin
model whose couplings can be tuned over a range of values thereby producing a
variety of interesting phases.
\end{abstract}

\maketitle

\section{Introduction}
\label{sec1}

Periodically driven quantum systems have been studied extensively over the last
several years, both theoretically~\cite{dunlap,gross,kaya,das,russo,nag,laza,
sharma,dutta,sengupta,agarwala1,agarwala2,lubini,frank,udupa,mukherjee1,
mukherjee2,haldar,eckardt1,rapp,zheng,greschner,laza1,alessio,ponte1,laza2,
ponte2,eckardt2,bukov,khemani,keyser,else,itin,mikami,su,raciunas,ghosh,mentink} and 
experimentally~\cite{bloch,kita,tarr,recht,mahesh,langen,jotzu,eckardt3,mciver,
meinert,bordia,mukh}
(see Refs.~\onlinecite{rev1,rev2,rev3,rev4,rev5,rev6,rev7} for reviews). Such 
systems can exhibit a wide variety interesting phenomena such as dynamical 
freezing~\cite{dunlap,das,nag,sengupta,agarwala1,agarwala2,lubini,mukherjee2,
haldar}, the generation of non-trivial band structures and states
localized at the boundaries of the system~\cite{oka,kitagawa10,lindner11,
jiang11,gu,trif12,gomez12,dora12,liu13,tong13,rudner13,katan13,lindner13,
kundu13,schmidt13,reynoso13,wu13,manisha13,perez,reichl14,carp,xiong,dutreix1,
manisha17,zhou,deb,sur}, and time crystals~\cite{khemani,else,zhang}.
While periodic driving of systems of non-interacting electrons has been studied 
very extensively, the effects of interactions along with periodic driving have
also been studied by several groups~\cite{eckardt1,rapp,zheng,greschner,laza1,
alessio,ponte1,laza2,ponte2,eckardt2,bukov,khemani,keyser,else,agarwala2,itin,
mikami,su,raciunas,sur,ghosh,meinert,bordia,mukh,mentink}. 

For undriven (time-independent) systems, it is often of interest to 
consider a subset of states such as the ground state and low-lying excitations 
which dominate the low-temperature properties of the system. For this purpose
it is convenient to find an effective low-energy Hamiltonian $H_{eff}$ which 
describes such states; the derivation of $H_{eff}$ usually involves taking 
into account all the other states in a perturbative way. For a system in 
which the Hamiltonian changes with time, we cannot define energy eigenstates
and there is no concept of low-energy states. However, as we will see,
we can define an effective time-independent Hamiltonian which describes
a particular sector of the system, such as the spin sector in which each site
is occupied by only one electron whose spin can point up or down.


In this paper, we will consider the Hubbard model of spin-1/2 electrons on a 
triangular lattice with a nearest-neighbor hopping amplitude $g$ and an 
on-site interaction $U$. In the absence of periodic driving, it is known that 
when the system is at half-filling and $U \gg g$, then up to order $g^2$, 
the low-energy Hamiltonian takes the 
form of a Heisenberg antiferromagnet with nearest-neighbor interactions of the 
form $J {\vec S}_i \cdot {\vec S}_j$ where $J = 4 g^2 /U$. We will consider
what happens when this model is subjected to a periodically varying electric
field which points in some direction in the plane of the lattice. The effect
of the electric field will be incorporated in the model using the Peierls
prescription~\cite{peierls}. 
We will show that up to order $g^3$, the Floquet Hamiltonian which
describes Floquet eigenstates in the spin sector (i.e., with large weights 
for states in which every site is singly occupied) has the following form. 
At order $g^2$ and $g^3$ there is a 
Heisenberg antiferromagnetic term $J_a {\vec S}_i \cdot {\vec S}_j$ which 
couples nearest neighbors but the coupling $J_a$ has three different values 
depending on the orientation of the bond joining the two sites. In addition, 
if the periodically varying electric field is not time-reversal symmetric,
a chiral three-spin interaction of the form $C {\vec S}_i \cdot {\vec S}_j 
\times {\vec S}_k$ can appear at order $g^3$ on each triangle, with $C$ having 
{\it opposite} values $\pm C$ on up- and down-pointing triangles. The values of
the four couplings, $J_a$ and $C$, can be tuned by varying the time-dependence
and direction of the periodically varying electric field and the driving
frequency $\om$. A spin model with four such couplings has not been 
studied earlier to the best of our knowledge although 
Refs.~\onlinecite{weng,yunoki,hauke,gonzalez} 
have studied models with different values of $J_a$ and 
$C=0$, and models with all $J_a$'s equal and $C$ having the {\it same} sign on
up- and down-pointing triangles have been studied in 
Refs.~\onlinecite{wietek} and \onlinecite{gong}.
We will then study the ground state phase diagram of our four-parameter spin
model using exact diagonalization (ED) of systems with
36 sites. We find a rich phase diagram consisting of three collinear ordered
phases, one coplanar ordered phase, and three disordered (spin-liquid) phases.
These phases can be distinguished from each other in several ways including
the locations of the peaks of the static spin structure function $S({\vec q})$ 
in the Brillouin zone of $\vec q$, the minimum value of the correlation function
$C({\vec r})$ in real space, the fidelity susceptibility of the ground 
state~\cite{rams}, and crossings of the energies of the ground state and first 
excited state.

The plan of the paper is as follows. In Sec.~\ref{sec2} we introduce the
Hubbard model on a triangular lattice in the presence of a periodically
varying electric field. In Sec.~\ref{sec3} we derive the effective spin Hamiltonian
using a Floquet perturbation theory which works in the limit that the
nearest-neighbor hopping is much smaller than the Hubbard interaction $U$ and
the driving frequency $\om$. We will find that there are nearest-neighbor 
Heisenberg antiferromagnetic terms with three different couplings $J_\al, ~
J_\beta, ~J_\ga$ and a chiral three-spin term with coefficient $\pm C$
on up- and down-pointing triangles. In Sec.~\ref{secc}, we study the
ground state phase diagram of the effective Floquet Hamiltonian in the 
classical limit and then use spin-wave theory to look at the excitations in 
the collinear phases. 
This is followed by Sec.~\ref{sec4} in which we study the spin model in detail
using ED for various values of $J_\al, ~J_\beta, ~J_\ga$ 
and $C$. We look at several quantities derived from the ground state wave 
function such as the static structure function in both real and momentum
space and the fidelity susceptibility to obtain the ground state phase
diagram. A rich phase diagram is found with four ordered phases and 
three spin-liquid phases. We conclude in Sec.~\ref{sec6} by summarizing our 
main results and pointing out possible directions for future studies.

\section{Hubbard model with periodic driving by electric field}
\label{sec2}

We consider the one-band Hubbard model of spin-1/2 electrons
on a triangular lattice at half-filling. The Hamiltonian is given by
\begin{equation} H ~=~ - ~g ~\sum_{<i,j>,\si} (c_{i,\si}^{\dg}
c_{j,\si} ~+~ {\rm H.c.}) ~+ ~U ~\sum_{i} ~n_{i,\ua}n_{i, \da},
\label{ham1} \end{equation}
where $g$ is the hopping amplitude between neighboring sites, and $U > 0$ is the 
on-site repulsive interaction. In the absence of driving and in the limit of large 
interaction, $ U \gg g$, the lowest energy sector of the Hamiltonian is 
described by an antiferromagnetic Heisenberg spin model at half-filling and 
by the \textit{t-J} model away from half-filling. Additionally, we drive 
the Hamiltonian periodically with a time varying in-plane electric field 
${\vec E}({\vec r},t) =  {\vec E}({\vec r},t+T)$.
To consider the most general cases, we will assume that the electric field is
{\it not} time-reversal symmetric, i.e., that there is no $t_0$ such that
$ {\vec E}({\vec r},t) =  {\vec E}({\vec r},t_{0}-t)$. This property of 
the electric field will turn out to be important for our study since, as we will
see, it gives rise to an additional term in the effective spin Hamiltonian. 
(Such an electric field can be realized by, say, superposing two sinusoidal electric 
fields with different frequencies and a phase difference as we will 
see in Sec.~\ref{sec4}). We will take the form of the electric field to be 
${\vec E} (t) = {\hat n} {\cal E} (t)$, where ${\hat n}$ denotes a unit vector
in the plane of the triangular lattice, and we will parametrize the direction of 
$\hat n$ by an angle $\theta$ with respect to the $\hat x$ axis. 

The time-dependent electric field is incorporated in our model
through a vector potential in the phase of the nearest-neighbor hopping 
following the Peierls prescription. Since ${\vec E} = - (1/c) 
\partial {\vec A}/\partial t$, the vector potential is ${\vec A}(t) = 
{\hat n} {\cal A} (t)$ where ${\cal A} (t)= - c \int_0^t dt' {\cal E} (t')$.
[We will assume that the electric field does not
have a dc component, i.e., $\int_0^T dt {\cal E} (t) = 0$. Then ${\cal A} (t)$ will
also be a periodic function of $t$.] The phase of the hopping from a site at ${\vec r}_j$ 
to a site at ${\vec r}_i$ is given by $(q/\hbar c) {\vec A} \cdot ({\vec r}_i - {\vec 
r}_j)$ where $q$ is the charge of the electron. Hence the periodic driving
modifies the term $- g c_{i,\si}^{\dg} c_{j,\si}$ in the Hamiltonian to 
$- g ~e^{i (q/\hbar c) {\cal A} (t) \hat{n}  \cdot ({\vec r}_i - {\vec r}_j)}~ 
c_{i,\si}^{\dg} c_{j,\si}$. 

\begin{figure}[H]
\centering
\includegraphics[width=0.5\textwidth]{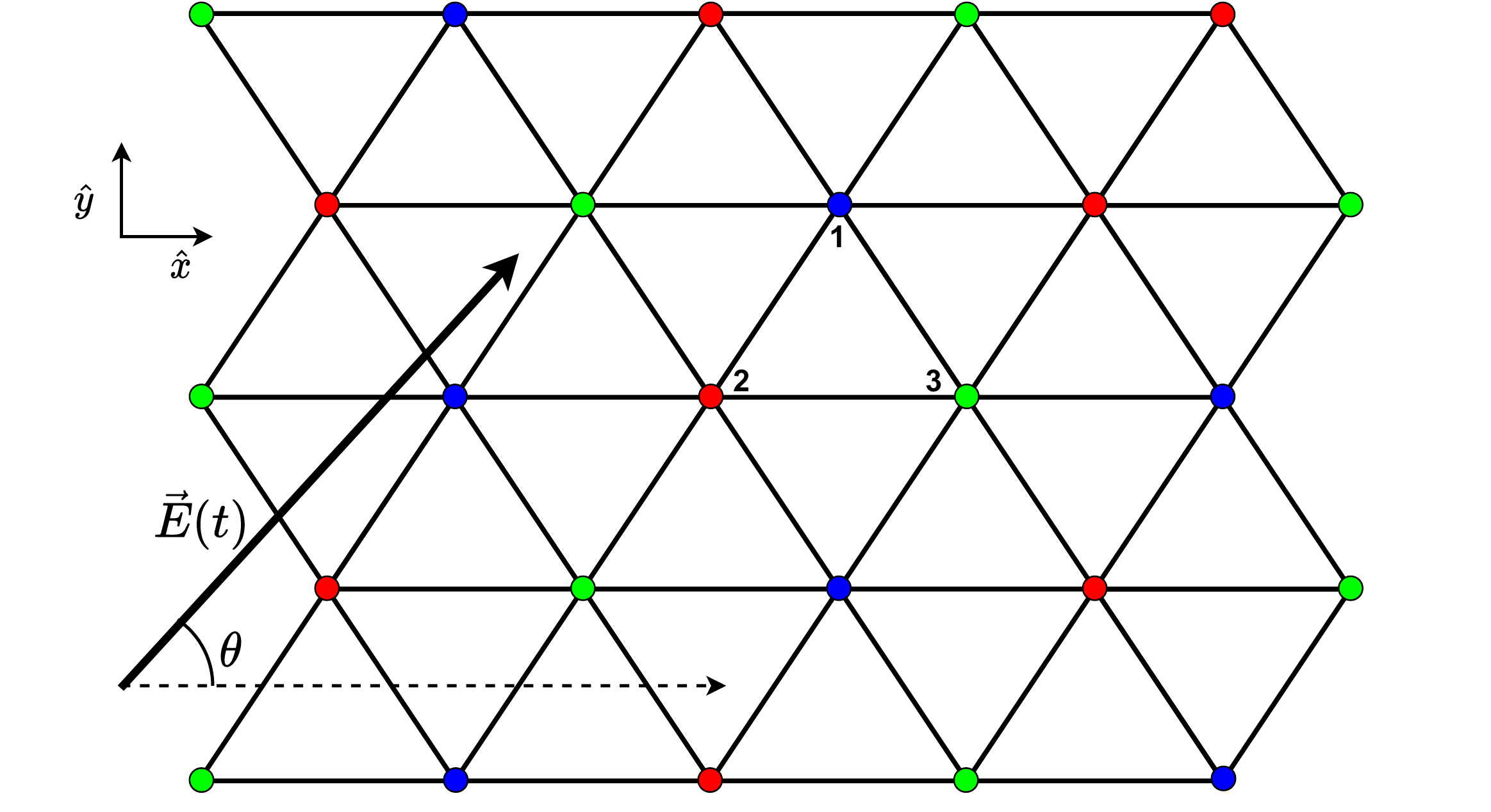}
\caption{A schematic of the triangular lattice model with the sublattices 
marked in three different colors. The sites are labeled in the anticlockwise 
direction $1$, $2$, and $3$ as shown. The time-dependent electric field 
$\vec{E}(t)$ points along a direction $\hat{n}$ which makes an angle $\theta$ with the 
$\hat{x}$ axis.} 
\label{fig:figure1}
\end{figure}

Figure~\ref{fig:figure1} illustrates how the different quantities look for a 
single triangle with sides of unit length whose sites are labeled as $1$, $2$ and 
$3$. (We have chosen the triangle to be up-pointing along the $\hat y$ axis.)
If $t_{ij}$ is the hopping amplitude from site-$j$ to site-$i$, we have
\begin{eqnarray}
t_{12} &=& g ~e^{~i (q/\hbar c) {\cal A} (t) \cos(\pi/3 - \theta)}, \non \\
t_{23} &=& g ~e^{~i (q/\hbar c) {\cal A} (t) \cos(\pi - \theta)}, \non \\
t_{31} &=& g ~e^{~i (q/ \hbar c) {\cal A} (t) \cos(\pi/3 + \theta)},
\label{hop1} \end{eqnarray}
and $t_{ji} ~=~ t^{*}_{ij}$. Then the periodically driven Hamiltonian for this 
triangle is 
\begin{eqnarray}
H_{\bigtriangleup}(t) &=& -\sum_{\sigma} (t_{12}~ c_{1,\si}^{\dg} c_{2,\si} + 
t_{23}~ c_{2,\si}^{\dg} c_{3,\si} + t_{31}~ c_{3,\si}^{\dg} c_{1,\si}  \non \\
&& ~~~~~~~~~+~ {\rm H.c.}) ~+ ~U ~\sum_{i = 1}^{3} ~n_{i,\ua}n_{i, \da}.
\label{ham2} \end{eqnarray}

To obtain the Hamiltonian for the entire triangular lattice, we sum up the 
Hamiltonians for all triangles, both up-pointing and down-pointing, with 
Hamiltonians $H_{\bigtriangleup}(t)$ and $H_{\bigtriangledown}(t)$.
In the next section we will use Floquet perturbation theory to derive the 
effective model in the large $U$ (Mott insulator) limit of this 
driven model. We shall see that the states in the spin sector (in which all
sites are occupied by only one electron) are governed by a 
Heisenberg Hamiltonian with different exchange couplings on different bonds and an 
additional chiral three-spin term whose sign is opposite for up-pointing 
($\bigtriangleup$) and down-pointing triangles ($\bigtriangledown$).

\section{Obtaining the effective spin Hamiltonian using Floquet perturbation theory}
\label{sec3}

To obtain the effective Hamiltonian in the large $U$ limit using Floquet perturbation theory, we start with the Hubbard model on a single triangle. We write the Hamiltonian in Eq. ~\eqref{ham2} as $H_{\bigtriangleup} ~= ~H_{0} + V$ where 
\begin{eqnarray}
H_{0} &=& ~U ~\sum_{i = 1}^{3} ~n_{i,\ua}n_{i, \da}, \non \\
V(t) &=& ~ -\sum_{\sigma} (t_{12}~ c_{1,\si}^{\dg} c_{2,\si} ~+~ t_{23}~ c_{2,\si}^{\dg} c_{3,\si} ~+~ t_{31}~ c_{3,\si}^{\dg} c_{1,\si}  \non \\
&& ~~~~~~~~~~+~ {\rm H.c.})   \non \end{eqnarray}
We now consider the eigenstates of the static part of the Hamiltonian, $H_{0}$, which will serve as the basis for subsequent calculations. For a half-filled system we can have three electrons on the triangle. Then the total number of basis states is $20$. They can be classified according to the number of up and down spins (in $S_{z}$ basis) which are
listed below:
\begin{itemize}
\item[(a)] One state with all three spins pointing up.
\item[(b)] Nine states with two up spins and a down spin. Six of these
states have a double occupancy.
\item[(c)] Nine states with one up and two down spins. Here also six states 
have a double occupancy.
\item[(d)] One state with all the three spins pointing down.
\end{itemize}
The reason for this classification is that these four sectors do not mix 
with each other since they have different values of the $z$-component of the
total spin, $S^z$, which commutes with both $H_0$ and $V(t)$. 

\begin{figure}[hb]
\centering
\includegraphics[width=0.5\textwidth]{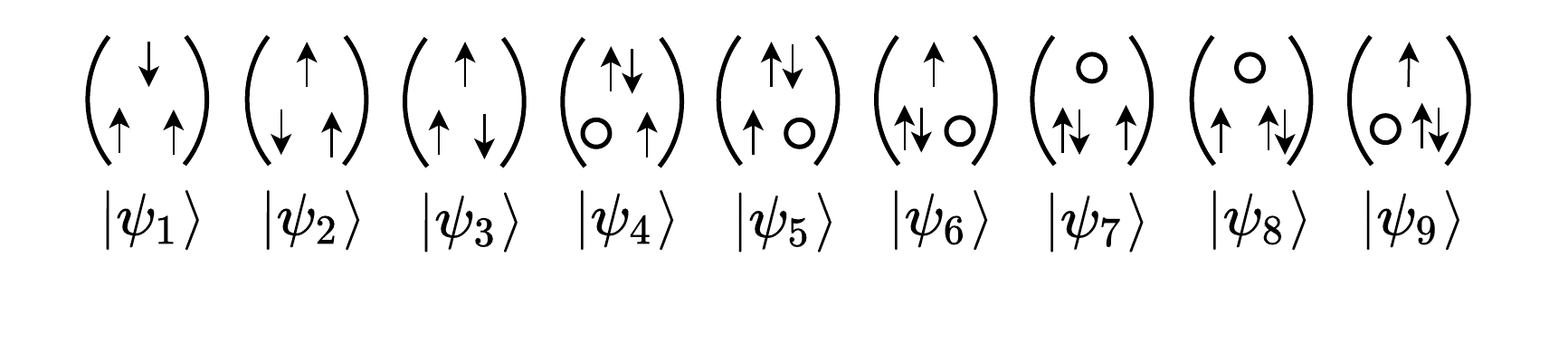}
\caption{The nine basis states in sector (b) are labeled from 
$\ket{\psi_{1}}$ to $\ket{\psi_{9}}$ as shown.} 
\label{fig:figure2}
\end{figure}

Sectors (a) and (d) are identical in terms of the 
eigenvalues of $H_0$ as are the sectors (b) and (c). Since the states in sector (a) and (d) are exact eigenstates of $V(t)$, the non-trivial eigenstates of $H_0$ are 
governed by states in sectors (b) and (c) which are related by the spin rotation 
operator which takes $S^z \rightarrow - S^z$. Hence we will derive the effective 
Hamiltonian for only sector (b). 
The nine basis states $\ket{\psi_{n}}$ in sector (b) are labeled as shown in 
Fig.~\ref{fig:figure2}. A general state $\ket{\psi(t)}$ with two up spins and one 
down spin can be written as a linear combination of these basis states as 
$\ket{\psi(t)}= \sum_{n=1}^{9} ~ c_{n}(t) \ket{\psi_n} ~e^{-iE_{n}t} $, where 
$E_{n}$'s are the eigenvalues of $H_{0}$. (We will set $\hbar $ equal to 1 in the 
rest of this paper). According to our notation $E_1 ~=~ E_2 ~=~ E_3 ~=~ 0$ and
$E_{n} ~=~ U$ for $n=4,5,\cdots,9$. We will follow the convention of defining 
the basis states in terms of the creation operators as follows. In 
Fig.~\ref{fig:figure2}, the state $\ket{\psi_1} ~=~ c_{1 
\downarrow }^{\dagger} c_{2 \uparrow }^{\dagger} c_{3 \uparrow }^{\dagger} \ket{0}$, 
whereas $\ket{\psi_4} ~=~ c_{1 \uparrow }^{\dagger} c_{1 \downarrow }^{\dagger} c_{3 
\uparrow }^{\dagger} \ket{0}$, namely, the three site labels are non-decreasing 
as we go from left to right, and at the same
site, $\uparrow$ appears to the left of $\downarrow$.

Next, $\ket{\psi(t)}$ satisfies the time-dependent Schr\"odinger equation $i \partial \ket{\psi(t)}/ \partial t = (H_{0} +V(t)) \ket{\psi(t)}$. Using the expansion of $\ket{\psi(t)}$, we obtain a set of linear differential equations for the coefficients $c_{n}$,
\begin{eqnarray}
i \frac{d c_{n}}{dt} = ~\sum_{m=1}^{9}~ \bra{\psi_n} V(t)\ket{\psi_m} ~ e^{i(E_{n}-E_{m})t} ~ c_m(t), \label{ceq1} \end{eqnarray}
where $n=1,2,\cdots,9$.

In our chosen basis, the matrix $\bra{\psi_n} V(t)\ket{\psi_m}$ matrix looks like 
\bea
\begin{pmatrix} 
0 & 0 & 0 & t_{21} & -t_{31} & 0 & t_{12} & -t_{13} & 0 \\
0 & 0 & 0 & -t_{21} & 0 & t_{32} & -t_{12} & 0 & t_{32} \\
0 & 0 & 0 & 0 & t_{31} & -t_{32} & 0 & t_{13} & -t_{23} \\
t_{12} & -t_{12} & 0 & 0 & -t_{32} & 0 & 0 & 0 & t_{13} \\
-t_{13} & 0 & t_{13} & -t_{23} & 0 & t_{12} & 0 & 0 & 0 \\
0 & t_{23} & -t_{23} & 0 & t_{21} & 0 & -t_{13} & 0 & 0 \\
t_{21} & -t_{21} & 0 & 0 & 0 & -t_{31} & 0 & t_{23} & 0 \\
-t_{31} & 0 & t_{31} & 0 & 0 & 0 & t_{32} & 0 & -t_{21} \\
0 & t_{32} & -t_{32} & t_{31} & 0 & 0 & 0 & -t_{12} & 0 \\
\end{pmatrix}. \non \\ 
\eea

Since the $t_{ij}$'s are periodic functions of time, they can be expanded as a 
Fourier series. The expressions for the time-dependent hoppings are therefore
\begin{eqnarray}
t_{12} &=& ~ g ~ \sum_{m=-\infty}^{\infty} \gamma_{m}~ e^{im\omega t}, \non \\
t_{23} &=& ~ g ~ \sum_{m=-\infty}^{\infty} \alpha_{m}~ e^{im\omega t}, \non \\
t_{31} &=& ~ g ~ \sum_{m=-\infty}^{\infty} \beta_{m}~ e^{im\omega t},
\label{hopf}
\end{eqnarray}
where $\alpha_{m}$, $\beta_{m} $ and $\gamma_{m}$ are generally complex. 

In Floquet theory, we define a Floquet operator which unitarily 
evolves the system through one time period $T$ as
\beq U_T ~=~ {\cal T} e^{-(i/\hbar) \int_0^T dt ~H(t)}, \label{ut} \eeq 
where $\cal T$ denotes time-ordering. If $\ket{\psi (t)}$ is an eigenstate of
$U_{T}$, we have 
\begin{equation} \ket{\psi (T)} = U_{T} \ket{\psi(0)}  = e ^{~-i \epsilon T} \ket{\psi (0)}, 
\label{floqeq1} \end{equation}
where $\epsilon$ is the quasienergy for the state $\ket{\psi(t)}$. Then in terms of the 
basis states $\ket{ \psi_n }$ and coefficients $c_n (t)$, we have
\begin{eqnarray}
c_{n}(T) ~ e^{~-i E_{n}T} &=& \sum_{m=1}^{9} \bra{\psi_{n}} U_{T}\ket{\psi_{m}} ~c_{m}(0)
\label{ceq2} \end{eqnarray}
for $n=1,2,\cdots,9$.
In the following subsections, we will solve Eq.~\eqref{ceq1} perturbatively in powers
of $g$ to obtain an effective Hamiltonian $H_{eff}$ such that $e^{~-i H_{eff}T} = U_T$ 
up to the desired power of $g$.

\subsection{Second-order calculation for effective Hamiltonian}
\label{sec3a}

Starting from Eq.~\eqref{ceq1} we solve for the coefficients of the low-energy states $c_{1}(T),~ c_{2}(T), ~c_{3}(T) $ up to order $g^{2}$ perturbatively. To begin with, the equations for $c_4(t) ~\cdots~ c_9(t) $ are solved at an arbitrary time $t$ assuming the coefficients $c_{1}(t), ~c_2 (t), ~c_{3}(t)$ to be constant and given by their
values at $t=0$. This is because $c_{4}(t) ~\cdots~ c_{9}(t)$ are coefficients 
of the states $\ket{\psi_4} ~\cdots ~\ket{\psi_9}$ which
lie in the high-energy sector and thus appear with coefficients which are of order $g$ times the coefficients of $c_{1}(t),~ c_2 (t), ~c_{3} (t)$. This procedure gives equations of the form
\begin{eqnarray}
c_{4}(t) - c_{4}(0) &=& -g\sum_{m} \gamma_{m} \frac{(e^{~i(U+m\omega)t}-1)}{U+m\omega}~ c_{1}(0) \non \\
&& + g\sum_{m} \gamma_{m} \frac{(e^{~i(U+m\omega)t}-1)}{U+m\omega}~ c_{2}(0), \non \\
\label{c4eq1}
\end{eqnarray}
and similar equations for $c_5(t)~ \cdots~ c_9(t) $. 
We now have to impose the Floquet condition Eq.~\eqref{floqeq1} for the coefficients. We have $c_{4}(T) ~=~ e^{~-i\epsilon T }e^{~iUT}~c_{4}(0) $. The Floquet eigenvalue $e^{~-i\epsilon T }$ gives the amplitude  $\braket{\psi(T)|\psi(0)}$, which is the
amplitude to start at time $t=0$ with a combination of the nine basis states
and come back to the same combination of states (up to overall phase) at time $t=T$. 
However this is either an order 1 or order $g^2$ process. This allows us to approximate $e^{~-i\epsilon T } = \mathbb{1} + O(g^{2})$. Hence, up to first order in $g$, we can write
\begin{equation}
c_{4}(T) - c_{4}(0) = (e^{~iUT} - 1)~c_{4}(0).
\end{equation}
Using this Floquet condition at $t=T$ in Eq. ~\eqref{c4eq1} we obtain the order $g$ expression for $c_{4}$
\begin{eqnarray}
c_{4}(t) &=& -g\sum_{m} \gamma_{m} \frac{e^{~i(U+m\omega)t}}{U+m\omega}~ c_{1}(0) \non \\
&& + ~g\sum_{m} \gamma_{m} \frac{e^{~i(U+m\omega)t}}{U+m\omega}~ c_{2}(0).
\label{c4eq2}                    
\end{eqnarray}
We get similar expressions for the coefficients $c_{5}(t) ~\cdots~ c_{9}(t)$.   

In the next step we substitute these $O(g)$ expressions for $c_{4}(t)~ \cdots~ c_{9}(t)$ in the right hand side of Eq.~\eqref{ceq1} to obtain the $O(g^{2})$ equations for $c_{1}(T), ~c_2 (T),~c_{3}(T)$, which eventually gives the second-order effective Hamiltonian for the Floquet system. We can write down final equations in a matrix form,
\begin{eqnarray}
\begin{pmatrix}
c_{1}(T) \\[0.2cm] 
c_{2}(T) \\[0.2cm]
c_{3}(T) 
\end{pmatrix}
~=~
(\mathbb{1} - iT H_{eff}^{(2)})~
\begin{pmatrix}
c_{1}(0) \\[0.2cm] 
c_{2}(0) \\[0.2cm]
c_{3}(0) 
\end{pmatrix}, \end{eqnarray}
where 
\bea H_{eff}^{(2)} &=& g^2 ~\begin{pmatrix}
- f_\beta - f_\ga & f_\ga & f_\beta \\
f_\ga & - f_\al - f_\ga & f_\al \\
f_\beta & f_\al & - f_\al - f_\beta \end{pmatrix}, \non \\
f_\al &=& \sum_{m} ~\abs{\alpha_{m}}^{2} ~( \frac{1}{U+m\omega} + \frac{1}{U-m\omega}), \non \\
f_\beta &=& \sum_{m} ~\abs{\beta_{m}}^{2} ~( \frac{1}{U+m\omega} + \frac{1}{U-m\omega}), 
\non \\
f_\ga &=& \sum_{m} ~\abs{\gamma_{m}}^{2} ~( \frac{1}{U+m\omega} + \frac{1}{U-m\omega}). \label{fabc} \eea
Comparing with Eq.~\eqref{ceq2} and noting that $E_{1}=E_{2}=E_{3}= 0$, we infer that ($\mathbb{1} - iT H_{eff}^{(2)}$) approximates $U_{T}$ to $O(g^{2})$.

We note that the quantities in Eq.~\eqref{fabc} diverge if $U/\omega$
approaches any integer values. This corresponds to a resonance condition,
and the coefficients $c_4 (t) ~\cdots~ c_9 (t)$ of the high-energy states 
will then not be much smaller than $c_{1}(t),~ c_{2}(t),~ c_{3}(t)$.
In our numerical calculations, we will choose $U$ and $\om$ in such
a way that $U/\om$ is not close to an integer.

\subsection{Third-order calculation for effective Hamiltonian}
\label{sec3b}

The third-order effective Hamiltonian is obtained by solving for $c_{1}(T),~ c_{2}(T),~ c_{3}(T) $ to $O(g^{3})$. Here we start with the $O(g)$ expressions for $c_{4}(t) \cdots~ c_{9}(t)$ which have already been calculated in Eq.~\eqref{c4eq2}. We now
use these expressions in the right hand side of the equations involving $c_{4}(t) ~\cdots~ c_{9}(t)$ in Eq.~\eqref{ceq1} to find the same expressions to the next order in $g$. 
We again use the Floquet condition to finally end up with the $O(g^{2})$ expressions for $c_{4}(t) ~\cdots~ c_{9}(t)$. The final expression for $c_4(t)$ in Eq.~\eqref{c4eq3} 
is given by
\begin{eqnarray}
c_{4}(t) &=& -g^{2} \sum_{m,n}~ \frac{\alpha_{n}^{*} \beta_{m}^{*}~ e^{~i(U-m\omega - n\omega)t}}{(U-m\omega)(U-m\omega - n\omega)} ~(c_{1}(0) - c_{3}(0)) \non \\
&& + g^{2} \sum_{m,n} ~\frac{\alpha_{m}^{*} \beta_{n}^{*}~ e^{~i(U-m\omega - n\omega)t}}{(U-m\omega)(U-m\omega - n\omega)} ~(c_{2}(0) - c_{3}(0)) .\non \\
\label{c4eq3}
\end{eqnarray}

Next, we use this expression along with similar expressions for $c_{5}(t) ~\cdots~ c_{9}(t)$ in the right hand side of Eq.~\eqref{ceq1} which finally gives us $O(g^{3})$ expressions for $c_{1}(T), ~c_2 (T), ~c_{3}(T)$. The third-order effective Hamiltonian $H_{eff}^{(3)}$ obtained from this calculation is given by
\begin{widetext}
\begin{eqnarray}
H_{eff}^{(3)} ~=~ g^{3}
\begin{pmatrix}
(d_{\alpha}+d_{\alpha}^{*}) - (e_{\alpha}+e_{\alpha}^{*}) & (e_{\alpha}^{*}+e_{\beta}^{*}+d_{\gamma}^{*})-(d_{\alpha}+ d_{\beta}+e_{\gamma})  & (e_{\alpha}+d_{\beta}+e_{\gamma})-(d_{\alpha}^{*}+ e_{\beta}^{*}+d_{\gamma}^{*}) \\[0.2cm] 
(e_{\alpha}+e_{\beta}+d_{\gamma})-(d_{\alpha}^{*}+ d_{\beta}^{*}+e_{\gamma}^{*}) & (d_{\beta}+d_{\beta}^{*}) - (e_{\beta}+e_{\beta}^{*})  & (d_{\alpha}^{*}+e_{\beta}^{*}+e_{\gamma}^{*})-(e_{\alpha}+ d_{\beta}+d_{\gamma}) \\[0.2cm]
(e_{\alpha}^{*}+d_{\beta}^{*}+e_{\gamma}^{*})-(d_{\alpha}+ e_{\beta}+d_{\gamma}) & (d_{\alpha}+e_{\beta}+e_{\gamma})-(e_{\alpha}^{*}+ d_{\beta}^{*}+d_{\gamma}^{*})  & (d_{\gamma}+d_{\gamma}^{*}) - (e_{\gamma}+e_{\gamma}^{*}) 
\end{pmatrix},
\non \\
\label{heff3} \end{eqnarray}
\end{widetext}
where we define 
\begin{eqnarray}
d_{\alpha} &=& \sum_{l,m} \frac{\alpha_{-(l+m)} ~\beta_{m} ~\gamma_{l}}{(U-m\omega)(U+l\omega)}, \non \\
d_{\beta} &=& \sum_{l,m} \frac{\beta_{-(l+m)} ~\gamma_{m} ~\alpha_{l}}{(U-m\omega)(U+l\omega)}, \non \\
d_{\gamma} &=& \sum_{l,m} \frac{\gamma_{-(l+m)} ~\alpha_{m} ~\beta_{l}}{(U-m\omega)(U+l\omega)}, \non \\
e_{\alpha} &=& \sum_{l,m} \frac{\alpha_{-(l+m)} ~\beta_{m} ~\gamma_{l}}{(U+m\omega)(U-l\omega)}, \non \\
e_{\beta} &=& \sum_{l,m} \frac{\beta_{-(l+m)} ~\gamma_{m} ~\alpha_{l}}{(U+m\omega)(U-l\omega)}, \non \\
e_{\gamma} &=& \sum_{l,m} \frac{\gamma_{-(l+m)} ~\alpha_{m} ~\beta_{l}}{(U+m\omega)(U-l\omega)}. \label{d-e-eqn} \end{eqnarray}

Thus, by considering the lowest energy states for a single triangle pointing upwards, 
we obtain the effective Hamiltonian for this driven system $ H_{\bigtriangleup}^{eff} = 
H_{eff}^{(2)} + H_{eff}^{(3)} $ up to  $O(g^{3})$. We can now rewrite this Hamiltonian 
in terms of spin operators at the three sites. Our original Hamiltonian Eq.~\eqref{ham1}, as 
well as the driven Hamiltonian Eq.~\eqref{ham2} are $SU(2)$ invariant, hence the effective 
Hamiltonian will also have the same spin rotational symmetry. The form of our Hamiltonian for 
three spin-1/2's on a triangle can therefore have only the following terms,
\begin{eqnarray}
H_{\bigtriangleup}^{eff} &=&  J_{\alpha}~ \vec{S_{2}} \cdot \vec{S_{3}} ~+~ J_{\beta} ~\vec{S_{3}} \cdot \vec{S_{1}} ~+~ J_{\gamma} ~\vec{S_{1}} \cdot \vec{S_{2}} \non \\
&& ~+~ C~ \vec{S_{1}} \cdot \vec{S_{2}} \times \vec{S_{3}} ~+~ D~ \mathbb{1}, \non \\
\label{Spinham} \end{eqnarray}
where $J_{\alpha}$, $J_{\beta}$, $J_{\gamma}$ are the two-spin exchange couplings, 
$C$ is a chiral three-spin term, and $D$ is a constant. 
In the basis states of sector (b), $(\ket{\psi_{1}}, \ket{\psi_{2}}, \ket{\psi_{3}})$, 
at sites 1, 2 and 3, this Hamiltonian has the following form:

\begin{widetext}
\begin{equation}
H_{\bigtriangleup}^{eff} ~\begin{pmatrix}
\ket{\psi_1} \\[0.2cm] \ket{\psi_2} \\[0.2cm] \ket{\psi_3}
\end{pmatrix} ~=~
\begin{pmatrix}
\frac{1}{4}(J_{\al}-J_{\beta}-J_{\ga})+D & \frac{1}{2} J_{\ga} + \frac{i}{4} C
& \frac{1}{2} J_{\beta} - \frac{i}{4} C \\[0.2cm]
\frac{1}{2} J_{\ga} - \frac{i}{4} C &
\frac{1}{4}(J_{\beta}-J_{\alpha}-J_{\ga})+D & 
\frac{1}{2} J_{\al} + \frac{i}{4} C \\[0.2cm]
\frac{1}{2} J_{\beta} + \frac{i}{4} C &
\frac{1}{2} J_{\al} - \frac{i}{4} C &
\frac{1}{4}(J_{\gamma}-J_{\alpha}-J_{\beta})+D
\end{pmatrix}
\begin{pmatrix}
\ket{\psi_1} \\[0.2cm] \ket{\psi_2} \\[0.2cm] \ket{\psi_3}
\end{pmatrix},
\label{hamS}    
\end{equation}
\end{widetext}

Comparing this equation with Eq.~\eqref{Spinham}, we obtain expressions for $J_{\al}$, $J_{\beta}$, $J_{\ga}$ and $C$. We further obtain the expression for $D$ using the special state of sector (a) with all three spins pointing up. This state has an eigenvalue equal to $\frac{1}{4}(J_{\al}+J_{\beta}+J_{\ga})+D$ for the Hamiltonian in Eq.~\eqref{Spinham}. But for our original Hamiltonian, this gives an energy eigenvalue equal to zero. Equating the two, we obtain $D$. The expressions for the five parameters are therefore given by
\begin{eqnarray}
J_\al &=& 2 g^2 f_\al ~-~ 2 g^{3} ~{\rm Re} ~[(d_{\alpha}+e_{\beta}+e_{\gamma})-(e_{\al} +d_{\beta} +d_{\gamma})], \non \\
J_\beta &=& 2 g^2 f_\beta ~-~ 2 g^3 ~{\rm Re} ~[(e_{\alpha}+d_{\beta}+e_{\gamma})-(d_{\al} +e_{\beta} +d_{\gamma})], \non \\
J_\ga &=& 2 g^2 f_\ga ~-~ 2 g^3 ~{\rm Re} ~[(e_{\al}+e_{\beta}+d_{\gamma})-
(d_{\al} +d_{\beta} + e_{\gamma})], \non \\
C&=& -4g^{3}~{\rm Im} ~[d_{\alpha}+d_{\beta}+d_{\gamma}+e_{\al} +e_{\beta} 
+e_{\gamma}], \non \\
D &=& -~\frac{1}{4}(J_{\al}+J_{\beta}+J_{\ga}). \label{forms} \end{eqnarray}

Interestingly, we observe that $C$ is zero when the $d$'s and $e$'s defined in Eq.~\eqref{d-e-eqn} are real. This is the case if the Hamiltonian is time-reversal symmetric, i.e., $H(t_{0}-t)=H^{\ast}(t)$ for some value of $t_{0}$. Then the Fourier expansions 
for the time-dependent hoppings obey $\sum_{m=-\infty}^{\infty} \alpha_{m}~ e^{im\omega(t_{0}- t)} = \sum_{m=-\infty}^{\infty} \alpha_{m}^{\ast}~ e^{-im\omega t}$. This
implies that $\alpha_{m}^{\ast} = \alpha_{m} ~ e^{im\omega t_{0}}$ for all $m$, and similarly for $\beta_{m}$'s and $\gamma_{m}$'s. This makes the $d$'s and $e$'s defined in Eq.~\eqref{d-e-eqn} completely real. 

However, for circularly polarized radiation where the vector potential is of
the form ${\vec A} (t) = A [\cos (\om t) {\hat x} + \sin (\om t) {\hat y}]$,
time-reversal symmetry is broken, but we find from Eqs.~\eqref{hop1}
and \eqref{forms} that $C$ vanishes at order $g^3$ for all values of $A$ 
and $\om$. (It turns out that we get a non-zero contribution to $C$ at
order $g^4$ as shown in Refs.~\onlinecite{claassen,kitamura,bostrom}).
Thus breaking time-reversal symmetry is a necessary but not sufficient 
condition to have a non-zero $C$ at order $g^3$.

The expressions for the second-order and third-order effective Hamiltonians
in Eqs.~\eqref{fabc} and \eqref{heff3} indicate that the perturbative 
expansion is valid if $g$ is much smaller than $U$, $\omega$ and 
$U + n \omega$ for any integer value of $n$. The condition that $g$ should be
much smaller than $U + n \omega$ for any $n$ is required to avoid resonances.

\subsection{Total effective Hamiltonian for the lattice}
\label{sec3c}

The effective Hamiltonian for a single up-pointing triangle, as derived above in the spin operator language, can be extended to the entire lattice. The important observation 
to note here is that the coefficient of the chiral three-spin 
term written in the the anticlockwise direction is opposite on up- and 
down-pointing triangles. [This is unlike the case of a time-independent 
magnetic field applied perpendicular to the plane of the lattice which gives 
the same sign of the chiral term for all triangles, both up- and 
down-pointing. This is because the chiral three-spin term then only depends 
on the magnetic flux through each triangle, and this has the same sign for all 
triangles~\cite{chitra}]. 
The reason for the sign flip in our model
is that when we go from $\bigtriangleup \rightarrow \bigtriangledown$, the angle that the external electric field makes with $\hat{x}$ changes as $\theta \rightarrow \pi + \theta$. This changes the hoppings as $t_{ij} \rightarrow t_{ij}^{\ast}$ and thus, $d_{k}, e_{k} 
\rightarrow d_{k}^{\ast}, e_{k}^{\ast}$. From Eq.~\eqref{forms} we see that this gives a negative sign on the right hand side in the expression for $C$. Hence $C \rightarrow -C$ when we go from an $\bigtriangleup$ triangle to a $\bigtriangledown$ triangle.

\begin{figure}[b!]
\centering
\includegraphics[width=0.5\textwidth]{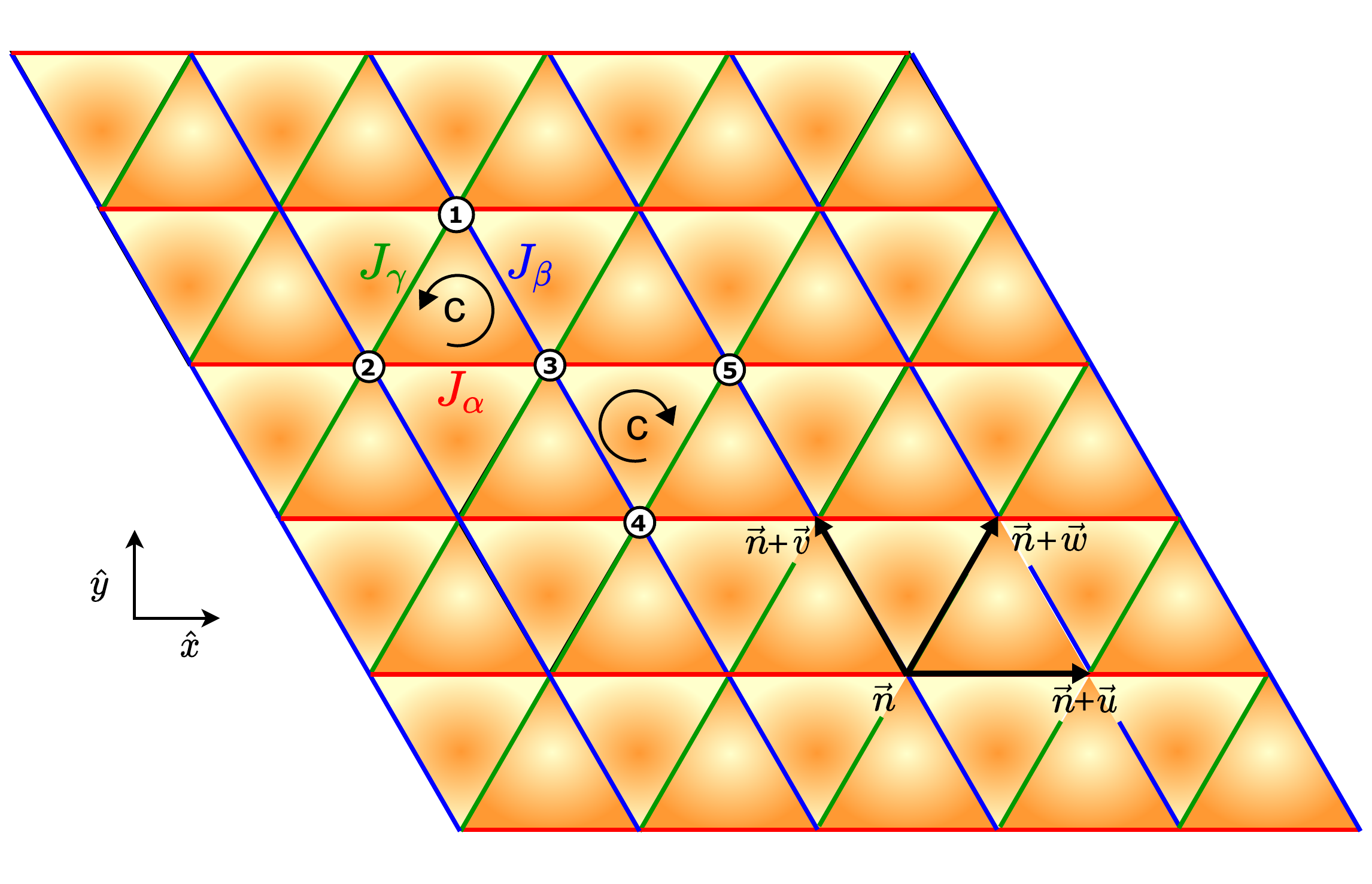}
\caption{A schematic of the triangular lattice model with the nearest-neighbor 
position vectors $\vu$, $\vv$ and $\vw$ and effective Heisenberg couplings $J_{\al}$, 
$J_{\beta}$ and $J_{\ga}$ for three different bonds as shown. The chiral three-spin 
term has opposite signs of $C$ for up-pointing (123) and down-pointing (345)
triangles.} 
\label{fig:figure3}
\end{figure}

The complete triangular lattice is made up of up-pointing and 
down-pointing triangles placed adjacently to each other. 
The total Hamiltonian for the lattice 
in the spin operator language can thus be written as
\begin{eqnarray}
H_L &=& \sum_{{\vec n}} ~[J_\al {\vec S}_{\vec n} \cdot {\vec S}_{\vec n 
+ \vec u} + J_\beta {\vec S}_{\vec n} \cdot {\vec S}_{\vec n + \vec v} + 
J_\ga {\vec S}_{\vec n} \cdot {\vec S}_{\vec n + \vec w}] \non \\
&& + ~C ~[\sum_{\vec n, \bigtriangleup} ~{\vec S}_{\vec n} \cdot 
{\vec S}_{\vec n + \vec u} \times {\vec S}_{\vec n + \vec w} \non \\
&& ~~~~~~~~~~-~ \sum_{\vec n, \bigtriangledown} {\vec S}_{\vec n} \cdot 
{\vec S}_{\vec n + \vec w} \times {\vec S}_{\vec n + \vec v}].
\end{eqnarray}
where $\vec{n}= m_1 \vu + m_2 \vv$ is the vector position of a site 
on the lattice, and
\bea \vec{u} &=& a ~\hat{x}, \non \\
\vec{v} &=& a~ (-\frac{1}{2} \hat{x} +\frac{\sqrt{3}}{2}\hat{y}), \non \\
\vec{w} &=& a~ (\frac{1}{2} \hat{x} + \frac{\sqrt{3}}{2}\hat{y}) 
\label{uvw} \eea 
as shown in Fig.~\ref{fig:figure3}. (We will henceforth set the 
nearest-neighbor lattice spacing $a=1$).  
The order of spin operators for the three-spin terms is conventionally taken to be in the anticlockwise direction. This convention gives a negative sign for $C$ for the down-pointing triangle. For instance, in Fig.~\ref{fig:figure3}, the three-spin term for the down-pointing triangle marked by sites $3$, $4$ and $5$ is $ ~-C~ \vec{S_{3}} \cdot \vec{S_{4}} \times \vec{S_{5}}$.

\section{Classical analysis and spin-wave theory}
\label{secc}

In this section we will consider the spin-$1/2$ Hamiltonian on the triangular 
lattice and study its classical limit where the value of the spin $S$ at each
site is taken to infinity. Having found the classical ground state we will
then use spin-wave theory to study the excitations above the classical 
ground state.

In the large-$S$ limit, we have to put a factor of $1/S$ in front of the
three-spin term so that it scales in the same way (i.e., as $S^2$) as 
the two-spin terms. We therefore consider the Hamiltonian 
\bea H_S &=& \sum_{{\vec n}} ~[J_\al {\vec S}_{\vec n} \cdot {\vec S}_{\vec n 
+ \vec u} + J_\beta {\vec S}_{\vec n} \cdot {\vec S}_{\vec n + \vec v} +
J_\ga {\vec S}_{\vec n} \cdot {\vec S}_{\vec n + \vec w}] \non \\
&& + ~\frac{C}{S} ~[\sum_{\vec n, \bigtriangleup} ~{\vec S}_{\vec n} \cdot 
{\vec S}_{\vec n + \vec u} \times {\vec S}_{\vec n + \vec w} \non \\
&& ~~~~~~~~~~- \sum_{\vec n, \bigtriangledown} ~{\vec S}_{\vec n} \cdot 
{\vec S}_{\vec n + \vec w} \times {\vec S}_{\vec n + \vec v}]. \label{hams} \eea
Since we are looking at the classical limit, we can take the 
components of spin to be commuting objects. We will now look at some
classical spin configurations and find the ranges of parameters 
$(J_\al,J_\beta,J_\ga,C)$ where each of these is stable.
Before proceeding further, we note that $\vec w = \vec u + \vec v$,
and the coordinates $\vec n$ of any site can be uniquely written as
\beq \vec n = m_1 \vec u + m_2 \vec v, \label{coord} \eeq
where $m_1, ~m_2$ are integers.

We first consider a collinear spin configuration in which all the 
spins point along the $\pm \hat z$ direction in the spin space, and they also satisfy 
\bea {\vec S}_{\vec n} \cdot {\vec S}_{\vec n + \vec u} &=& - S^2,
\non \\ 
{\vec S}_{\vec n} \cdot {\vec S}_{\vec n + \vec v} &=& S^2, \non \\
{\vec S}_{\vec n} \cdot {\vec S}_{\vec n + \vec w} &=& - S^2,
\label{spinconf1} \eea
for all values of $\vec n$. Following Fig.~\ref{fig:figure3} and
Eq.~\eqref{coord}, we see that a spin configuration which satisfies
Eqs.~\eqref{spinconf1} is given by
\beq {\vec S}_{\vec n} ~=~ S (0, 0, (-1)^{m_1}). \label{spinconf2} \eeq
For a large system with $N$ sites and periodic boundary conditions, we 
find that the classical energy of this configuration is given by
\beq E_{cl} ~=~ N S^2 ~(- ~J_\al ~+~ J_\beta ~-~ J_\ga). \label{classen1} \eeq
[Note that the three-spin term in Eq.~\eqref{hams} does not contribute 
to the energy in any collinear spin configuration. We will therefore set $C=0$
in the rest of this analysis.] We will now use
spin-wave theory~\cite{anderson,kubo} to find the energy-momentum dispersion 
of the excitations around the classical ground state in Eq.~\eqref{spinconf2}.
We use the Holstein-Primakoff transformation~\cite{holstein} to write the 
spin operators in terms of bosonic operators as
\bea S_\vn^z &=& S ~-~ {a_\vn}^\dg a_\vn, \non \\
S_{\vec n}^+ &=& \sqrt{2S ~-~ a_{\vec n}^\dg a_{\vec n}} ~a_{\vec n} \non \\
S_{\vec n}^- &=& a_{\vec n}^\dg ~\sqrt{2S ~-~ a_{\vec n}^\dg a_{\vec n}}, \label{hols1} \eea
at the sites where ${\vec S}_\vn = S (0,0,1)$, and 
\bea S_{\vec n}^z &=& -~ S ~+~ a_{\vec n}^\dg a_{\vec n}, \non \\
S_{\vec n}^+ &=& a_{\vec n}^\dg \sqrt{2S ~-~ a_{\vec n}^\dg a_{\vec n}}, \non \\
S_{\vec n}^- &=& \sqrt{2S ~-~ a_{\vec n}^\dg a_{\vec n}} ~a_{\vec n}, \label{hols2} \eea
at the sites where ${\vec S}_\vn = S (0,0,-1)$. Making the standard large-$S$ approximation
of replacing $\sqrt{2S ~-~ a_{\vec n}^\dg a_{\vec n}} \to \sqrt{2S}$, we find that 
Eq.~\eqref{hams} takes the form
\bea H_S &=& N S^2 ~(- ~J_\al ~+~ J_\beta ~-~ J_\ga) \non \\
&& +~ S ~\sum_\vn ~[(2J_\al ~-~ 2J_\beta ~+~ 2J_\ga) ~a_\vn^\dg a_\vn \non \\
&& ~~~~~~~~~~~~~~~ +~ J_\al ~(a_\vn^\dg a_{\vn + \vu}^\dg ~+~ {\rm H.c.}) \non \\
&& ~~~~~~~~~~~~~~~ +~ J_\beta ~(a_\vn^\dg a_{\vn + \vv} ~+~ {\rm H.c.}) \non \\
&& ~~~~~~~~~~~~~~~ +~ J_\ga ~(a_\vn^\dg a_{\vn + \vw}^\dg ~+~ {\rm H.c.})]. 
\label{hams2} \eea
Fourier transforming to momentum space, we obtain
\bea 
H_S &=& N S^2 ~(- ~J_\al ~+~ J_\beta ~-~ J_\ga) \non \\
&& +~ S ~\sum_\vk ~[(2J_\al ~-~ 2J_\beta ~+~ 2J_\ga) ~a_\vk^\dg a_\vk \non \\
&& ~~~~~~~~~~~~~~~ +~ J_\al ~\cos (\vk \cdot \vu) ~(a_\vk^\dg a_{-\vk}^\dg ~+~ 
a_\vk a_{- \vk}) \non \\
&& ~~~~~~~~~~~~~~~ +~ 2 J_\beta ~\cos (\vk \cdot \vv) ~a_\vk^\dg a_\vk \non \\
&& ~~~~~~~~~~~~~~~ +~ J_\ga ~\cos (\vk \cdot \vw) ~(a_\vk^\dg a_{-\vk}^\dg ~+~ 
a_\vk a_{- \vk})], \non \\
\label{hams3} \eea
where the sum over $\vk$ runs over the complete Brillouin zone. The above
Hamiltonian couples modes at $\vk$ and $- \vk$. Using the Bogoliubov 
transformation to diagonalize the Hamiltonian, we find that the spin-wave
spectrum is given by
\bea E_\vk &=& S~ \sqrt{(C_\vk ~+~ D_\vk) ~(C_\vk ~-~ D_\vk)}, \non \\
C_\vk &=& J_\al + J_\ga - J_\beta + J_\beta ~\cos (\vk \cdot \vv), \non \\
D_\vk &=& J_\al ~\cos (\vk \cdot \vu) ~+~ J_\ga ~\cos (\vk \cdot \vw). 
\label{disp1} \eea
We see that the spin-wave energy vanishes at $\vk = (0,0)$ and 
$\vk = \pi (1,1/\sqrt{3})$ which correspond to Goldstone modes.
We therefore expect that in this phase the static structure function 
$S(\vq)$ should have a peak at $\vq = \pi (1,1/\sqrt{3})$.
We can also see this directly from the form of the classical spin
configuration in Eq.~\eqref{spinconf2}. Since the two-spin
correlation between sites ${\vec 0}=(0,0)$ and $\vn$ is equal to 
${\vec S}_{\vec 0} \cdot {\vec S}_{\vn} = S^2 (-1)^{m_1}$, we see 
from Eqs.~\eqref{uvw} and \eqref{coord} that the Fourier transform,
\beq S(\vq) ~=~ \sum_\vn ~e^{-i \vq \cdot \vn} ~{\vec S}_{\vec 0} 
\cdot {\vec S}_{\vn}, \label{sq1} \eeq
will have a peak at $\vq = \pi (1,1/\sqrt{3})$.
We will see later that this agrees with our numerical results based
on ED.

Expanding around $\vk = (0,0)$, we find that
\bea E_\vk^2 &=& S^2 a^2 ~(J_\al + J_\ga) \left( \begin{array}{cc}
k_x & k_y \\
\end{array} \right) ~M_\vk~ \left( \begin{array}{c}
k_x \\
k_y \end{array} \right), \non \\
M &=& \left( \begin{array}{cc}
J_\al + \frac{1}{4} (J_\ga - J_\beta) & \frac{\sqrt 3}{4} (J_\beta + J_\ga) \\
\frac{\sqrt 3}{4} (J_\beta + J_\ga) & \frac{3}{4} (J_\ga - J_\beta) \end{array} 
\right). \label{disp2} \eea
The above analysis clearly breaks down if any of the eigenvalues of $M_\vk$
becomes negative since that would make the energy $E_\vk$ imaginary. This
happens if
\beq {\rm det}~ (M_\vk) ~=~ \frac{3}{4} ~(J_\al J_\ga ~-~ J_\beta J_\ga 
~-~ J_\al J_\beta) \label{detm} \eeq
turns negative. We thus conclude that the spin-wave spectrum near the 
ground state spin configuration given in Eq.~\eqref{spinconf1} is real if
\beq \frac{1}{J_\beta} ~>~ \frac{1}{J_\al} ~+~ \frac{1}{J_\ga}, \label{ph1} \eeq
and a transition must occur to some other phase when
\beq \frac{1}{J_\beta} ~=~ \frac{1}{J_\al} ~+~ \frac{1}{J_\ga}. \label{pt1} \eeq
As we approach the line in Eq.~\eqref{pt1} from the region in Eq.~\eqref{ph1},
we find from Eq.~\eqref{disp2} that one of the spin-wave energies remains finite
while the other approaches zero as some finite constant times $\sqrt{\lm} |\vk|$, where
\beq \lm ~\equiv~ \frac{1}{J_\beta} ~-~ \frac{1}{J_\al} ~+~ \frac{1}{J_\ga}.
\label{lamb} \eeq

The above analysis was based on an expansion of the spin Hamiltonian in 
Eq.~\eqref{hams} up to order $S$, assuming that the ground state expectation
value of $a_\vn^\dg a_\vn$ appearing in Eqs.~\eqref{hols1} and \eqref{hols2}) 
are much smaller than $S$. We can now check for the self-consistency of this
assumption~\cite{anderson}. We find that
\beq \langle a_\vn^\dg a_\vn \rangle ~\sim~ \int \frac{d^2 k}{E_\vk}. 
\label{ada} \eeq

Near the vicinity of the line in Eq.~\eqref{pt1} and $\vk = (0,0)$, we see from
Eq.~\eqref{lamb} that the integral in Eq.~\eqref{ada} diverges as $1/\sqrt{\lm}$. Hence, for
large $S$, the spin-wave analysis is expected to break down in a region appearing
{\it before} the phase transition line where the integral in Eq.~\eqref{ada} is not much smaller
than $S$. We may expect this region to form a disordered phase. We will see in 
the next section that the ground state phase diagram for our model with $S=1/2$ 
indeed has some disordered phases lying between the ordered phases.

\begin{figure}[htb]
\centering
\includegraphics[width=0.45\textwidth]{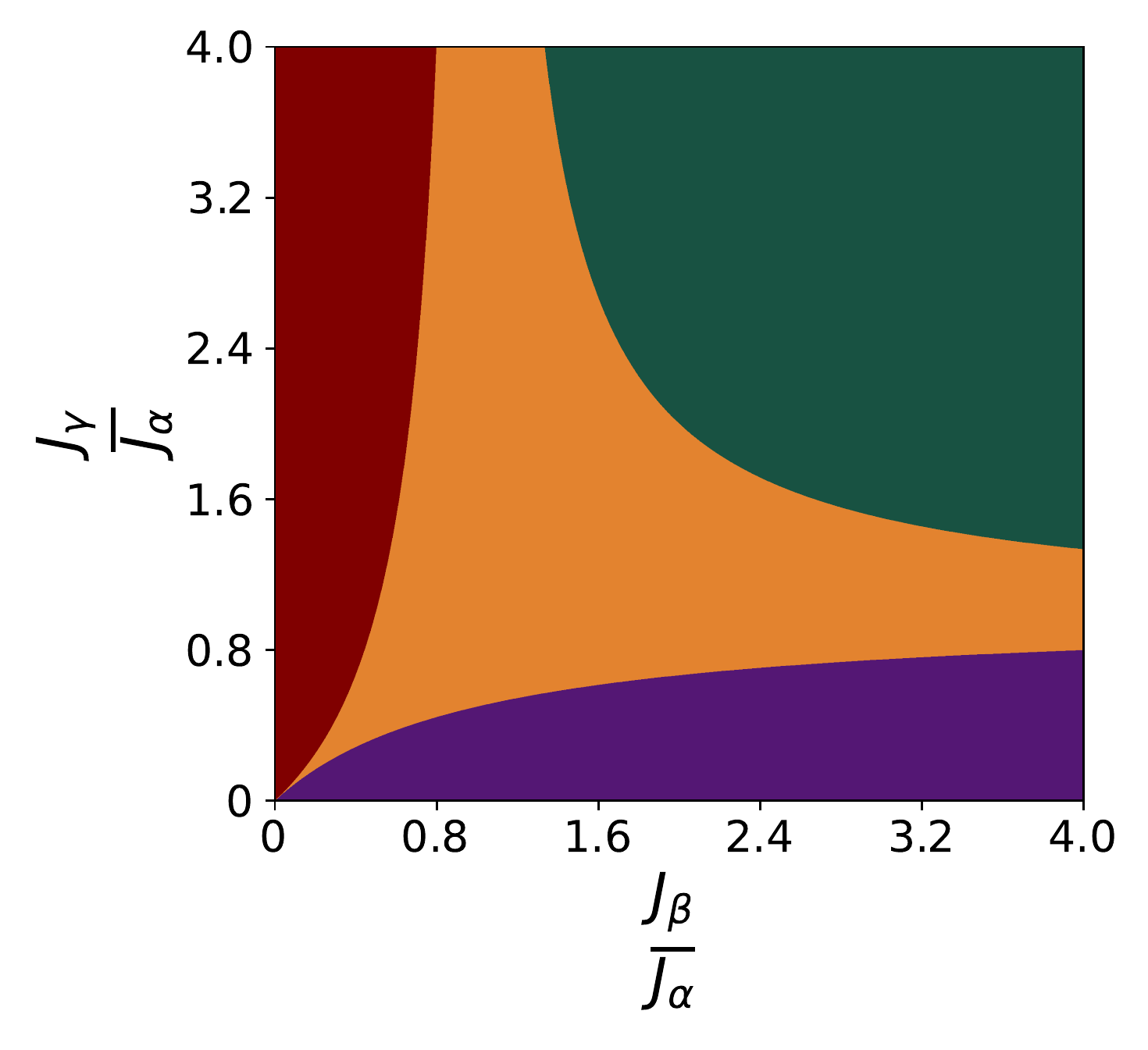}
\caption{The classical ground state phase diagram as a function 
of $J_\beta/J_\al$ and $J_\ga /J_\al$ for $C=0$. There are three 
collinear (or striped) phases where Eqs.~\eqref{ph1} (shown in maroon),
\eqref{ph2} (dark green) and \eqref{ph3} (violet) are satisfied. The
rest of the diagram, shown in yellow, describes a coplanar (or spiral)
phase.} \label{fig:figure4} \end{figure}

By permuting between the three couplings $J_\al, ~J_\beta$ and $J_\ga$, we find 
that there must be two other regions similar to Eq.~\eqref{ph1}, namely, 
\beq \frac{1}{J_\al} ~>~ \frac{1}{J_\beta} ~+~ \frac{1}{J_\ga} \label{ph2} \eeq
and
\beq \frac{1}{J_\ga} ~>~ \frac{1}{J_\al} ~+~ \frac{1}{J_\beta}, \label{ph3} \eeq
where a possible ground state spin configuration is given by 
\beq {\vec S}_{\vec n} ~=~ S (0, 0, (-1)^{m_2}) \label{spinconf3} \eeq
and
\beq {\vec S}_{\vec n} ~=~ S (0, 0, (-1)^{m_1+m_2}) \label{spinconf4} \eeq
respectively. The three collinear ordered phases given by Eqs.~\eqref{ph1}, 
\eqref{ph2} and \eqref{ph3} are shown in Fig.~\ref{fig:figure4} (see 
also Ref.~\onlinecite{hauke}).

We will now briefly discuss the remaining region in Fig.~\ref{fig:figure4}, still setting
$C=0$. We will assume that the ground state spin configuration in this region is
given by a coplanar configuration of the form
\bea {\vec S}_\vn &=& S ~(\cos \phi_\vn, \sin \phi_\vn, 0), \non \\
{\rm where} ~~~\phi_\vn &=& m_1 \phi_1 + m_2 \phi_2, \label{spinconf5} \eea
for $\vn = m_1 \vu + m_2 \vv$. This implies that
\bea {\vec S}_{\vec n} \cdot {\vec S}_{\vec n + \vec u} &=& S^2 ~\cos \phi_1,
\non \\ 
{\vec S}_{\vec n} \cdot {\vec S}_{\vec n + \vec v} &=& S^2 ~\cos \phi_2, \non \\
{\vec S}_{\vec n} \cdot {\vec S}_{\vec n + \vec w} &=& S^2 ~\cos (\phi_1 + \phi_2).
\label{spinconf6} \eea
The classical ground state energy for a system with $N$ sites is then given by
\beq E_{cl} ~=~ N S^2 ~[J_\al ~\cos \phi_1 ~+~ J_\beta ~\cos \phi_2 ~+~ 
J_\ga ~\cos (\phi_1 + \phi_2)]. \label{classen2} \eeq
Minimizing Eq.~\eqref{classen2} with respect to the angles $\phi_1, ~\phi_2$, we
obtain
\bea J_\al ~\sin \phi_1 ~+~ J_\ga ~\sin (\phi_1 + \phi_2) &=& 0, \non \\
J_\beta ~\sin \phi_2 ~+~ J_\ga ~\sin (\phi_1 + \phi_2) &=& 0. \label{phi12} \eea
Given some values of $J_\al, J_\beta$ and $J_\ga$, we can numerically solve 
Eqs.~\eqref{phi12} for 
$\phi_1, ~\phi_2$ to find a classical ground state configuration.
We will not discuss here the spin-wave theory about such a ground state.
For $J_\al = J_\beta = J_\ga$, there are two solutions of Eqs.~\eqref{phi12} given
by $\phi_1 = \phi_2 = 2 \pi/3$ and $\phi_1 = \phi_2 = 4 \pi/3$. We also note that 
the collinear spin configurations given in Eqs.~\eqref{spinconf2}, \eqref{spinconf3} 
and \eqref{spinconf4} are, after doing a rotation which transforms the $\hat x$
axis into the $\hat z$ axis, special cases of Eq.~\eqref{spinconf5} corresponding 
to $(\phi_1,\phi_2) = (\pi,0)$, $(0,\pi)$ and $(\pi,\pi)$ respectively.

For the spin configuration given in Eq.~\eqref{spinconf5}, we have 
${\vec S}_{\vec 0} \cdot {\vec S}_\vn = S^2 \cos (m_1 \phi_1 + m_2 \phi_2)$.
Then an argument similar to the one around Eq.~\eqref{sq1} implies that
$S (\vq)$ will have $\de$-function peaks at $\vq = \pm (\phi_1, (\phi_1 + 
2 \phi_2)/\sqrt{3})$. Equation~\eqref{phi12} implies that the locations of the peaks
will move around as the parameters $J_\al, ~J_\beta$ and $J_\ga$ are changed. 
In contrast to this, we 
will see later that our numerical results show peaks which are fixed at 
$\vq = \pm (2 \pi/3, 2 \pi/\sqrt{3})$ for all points in the coplanar phase.
This is a qualitative difference between the classical (large $S$) and quantum
($S=1/2$) models.

Finally, we consider the case $C \ne 0$ and equal two-spin interactions, 
$J_\al = J_\beta = J_\ga$. Now we find that the classical ground state 
configuration is neither collinear nor coplanar. We assume that the ground state spin
configuration is given by
\bea {\vec S}_\vn &=& S ~(\sin \ta \cos \phi_\vn, \sin \ta \sin \phi_\vn, \cos \ta), 
\non \\
{\rm where} ~~~\phi_\vn &=& (m_1 + m_2)~ \frac{2\pi}{3}, \label{spinconf7} \eea
for $\vn = m_1 \vu + m_2 \vv$. This implies that
\bea {\vec S}_{\vec n} \cdot {\vec S}_{\vec n + \vec u} &=& {\vec S}_{\vec n} 
\cdot {\vec S}_{\vec n + \vec v} ~=~ {\vec S}_{\vec n} \cdot {\vec S}_{\vec n + 
\vec w} \non \\
&=& S^2 ~(\cos^2 \ta ~-~ \frac{1}{2} ~\sin^2 \ta), \label{snuvw1} \\
{\vec S}_{\vec n} \cdot {\vec S}_{\vec n + \vec u} \times {\vec S}_{\vec n + \vec w} 
&=& {\vec S}_{\vec n} \cdot {\vec S}_{\vec n + \vec v} \times 
{\vec S}_{\vec n + \vec w} \non \\
&=& S^3 ~\frac{3\sqrt{3}}{2} ~\cos \ta ~\sin^2 \ta. \label{snuvw2} \eea
For a system with $N$ sites, the ground state energy is then
\beq E_{cl} ~=~ N S^2 ~[3J_\al ~(\cos^2 \ta ~-~ \frac{1}{2} ~\sin^2 \ta) ~+~ 
3\sqrt{3} C ~\cos \ta ~\sin^2 \ta], \label{classen3} \eeq
where we have used the fact that for each site, there is one up-pointing and one
down-pointing triangle. (It is important to note here that this calculation
works only because the coefficient $C$ of the three-spin term in Eq.~\eqref{hams}
has opposite signs for the two kinds of triangles; if the sign had been the same
for all triangles, the analysis of the classical ground state spin configuration
would have been significantly more complicated). Minimizing Eq.~\eqref{classen3} 
with respect to $\ta$, we obtain
\beq \cos \ta ~=~ \frac{1}{2 \sqrt{3} C} ~[J_\al ~-~ \sqrt{J_\al^2 ~+~ 4 C^2}].
\eeq
We find that for $C=0$, $\ta = \pi/2$ which agrees with the discussion in the 
previous paragraph (with $\phi_1 = \phi_2 = 2 \pi/3$); we thus recover a coplanar 
spin configuration with the expressions in Eqs.~\eqref{snuvw1} and \eqref{snuvw2} 
being equal to $-S^2/2$ and zero respectively. 
For $C/J_\al \to \pm \infty$, $\ta \to \arccos (-{\rm sgn}
(C)/\sqrt{3})$, where ${\rm sgn} (C) = + 1 ~(-1)$ for $C > 0 ~(< 0)$. Hence
Eqs.~\eqref{snuvw1} and \eqref{snuvw2} are equal to zero and 
$-{\rm sgn} (C) ~S^3$ respectively, i.e., in each triangle the three spins
are perpendicular to each other.

\section{Numerical analysis of the model}
\label{sec4}

Having derived the lattice Hamiltonian for our model, we will now do an ED 
study to look at the ground state properties as a function of the parameters $(J_\al,J_\beta,J_\ga,C)$. The triangular lattice is spanned by the primitive unit cell vectors $\vec{u}$ and $\vec{v}$ as shown in Fig.~\ref{fig:figure3}. We choose to perform our ED calculations on a $6 \times 6$ lattice system with total number of lattice sites, $N=36$ with periodic boundary conditions applied in both the directions. This system size is particularly well suited for our purpose since this ensures there is no frustration in the sublattice symmetry of the triangular lattice in both the directions and the number of
spin-1/2's is even. This enables us to work in the zero magnetization sector for the ground state calculations. We make use of the following symmetries in the system: (i) translation along $\hat{u}$ direction, (ii) translation along $\hat{v}$ direction, (iii) total magnetization $m$ in the $\hat z$-direction in spin space, and (iv) spin inversion by
the operator $Z= e^{i\pi \Pi_{\vec{n}}S_{\vec{n}}^{x}}$ which flips $S_\vn^z \to - S_\vn$ at every site (with $Z=1$ for the even sector and $Z=-1$ for the odd sector). 

For the ground state, we work in the momentum sector $(q_x, q_y)= (0,0)$, zero magnetization sector $m=0$, and even spin inversion sector with eigenvalue of $Z=1$. In addition, for the case when $C=0$, we also have (v) simultaneous spatial inversion 
symmetry $P$ along the $\hat{u}$ and $\hat{v}$ directions. The operator $P=P_xP_y$ acting on the state takes $x \rightarrow L_x-x$ and $y \rightarrow L_y-y$, where $L_x$ 
and $L_y$ are the lengths of the system along $\vec{u}$ and $\vec{v}$ respectively. The ground state has an even parity for this operator enabling us to do the diagonalization 
in this sector. The use of these symmetries reduces the Hilbert space dimension from $2^{36}$ (about $6.8 \times 10^{10}$) down to about
$6.3 \times 10^{7} $. 
We then examine in detail at the spatial correlation function, static spin structure function (SSSF), and fidelity susceptibility for the ground state as a function 
of the parameters $J_{\al}$, $J_{\beta}$, $J_{\gamma}$, and $C$. 

For our numerical studies, we will consider an electric field which does not have
time-reversal symmetry. As an example, we will take the electric field to be
\beq {\vec E} (t) ~=~ {\hat n} ~[{\cal E}_1 \cos (\omega t) + {\cal E}_2 
\sin (2\omega t)]. \eeq
We note that this electric field is not time-reversal symmetric unless 
${\cal E}_2 = 0$. Following the steps leading up to Eq.~\eqref{hop1}, we now
find that the hopping amplitudes are given by
\begin{eqnarray}
t_{12} &=& ~ g ~ e^{~(i/\omega) ~ [a_1 \sin (\omega t) + (a_2/2)  
\cos (2\omega t)]~ \cos(\pi/3 -\theta)}, \non \\
t_{23} &=& ~ g ~ e^{~(i/\omega) ~ [a_1 \sin (\omega t) + (a_2/2)
\cos (2\omega t)]~ \cos(\pi -\theta)}, \non \\
t_{31} &=& ~ g ~ e^{~(i/\omega) ~ [a_1 \sin (\omega t) + (a_2/2) 
\cos (2\omega t)]~ \cos (\pi/3 + \theta)}, \end{eqnarray}
where $a_1 = - q{\cal E}_1/ \hbar$ and $a_2 = q {\cal E}_2/ \hbar$, 
and $t_{ji}=t^*_{ij}$.

\subsection{Numerical values of $J_{\alpha}$, $J_{\beta}$, $J_{\gamma}$ 
and $C$ from periodic driving}
\label{sec5a}

\begin{figure}[htb]
\centering
\includegraphics[width=0.5\textwidth]{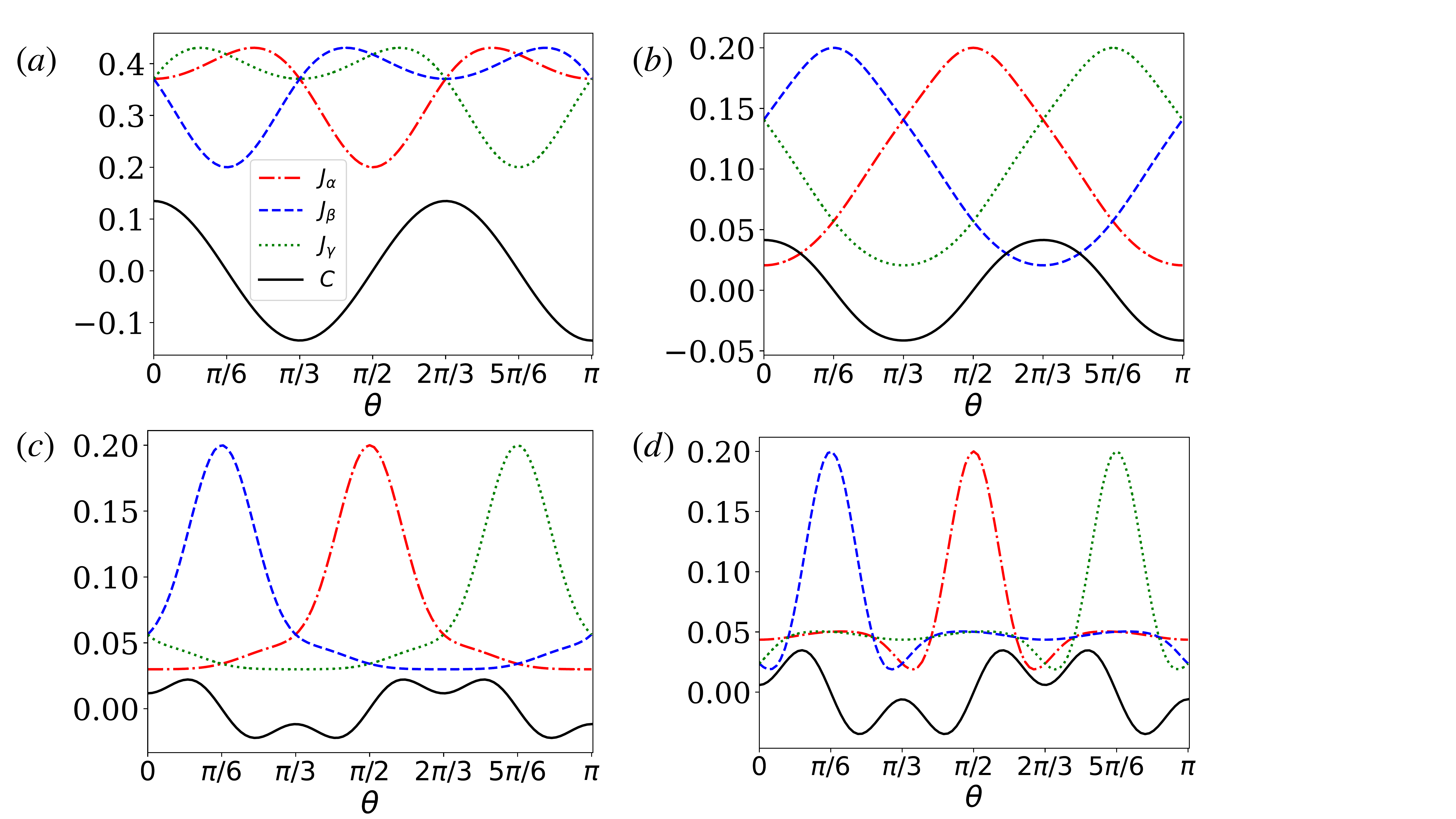}
\caption{Plots of $J_{\al}$, $J_{\beta}$, $J_{\gamma}$ and $C$ as functions of $\theta$, for $g=1$, $U=20$, $\omega=17$, and $a_1 = 35$.
We have taken (a) $a_2$= 30, 
(b) $a_2$= 105, (c) $a_2$= 140, and (d) $a_2$= 175. The values of 
$a_2$ have been chosen so as to show four qualitatively different 
behaviors versus $\ta$.} \label{fig:figure5} \end{figure}

The four parameters $J_{\al}$, $J_{\beta}$, $J_{\gamma}$ and $C$ depend on the amplitudes of driving $a_1$ and $a_2$, the 
frequency of driving $\omega$, the direction of the electric field 
$\theta$, and the interaction strength $U$. 
We will set $g=1$ in all the numerical calculations. 
A simple parameter to vary in an experimental set-up would be the 
electric field direction. Hence, we first fix the values of $a_1$, $a_2$, $\omega$ and $U$, and look at the variation of the $J_{\al}$, $J_{\beta}$, $J_{\gamma}$ and $C$ with $\theta$. Figure~\ref{fig:figure5} shows the plots
of these parameters obtained using the expressions from third-order
perturbation theory given in Eq.~\eqref{forms}.
We notice that the couplings $J_{\al}$, $J_{\beta}$ and $J_{\gamma}$ vary with a periodicity of $\pi$. Further,
the values of $J_{\al}$, $J_{\beta}$ and $J_{\gamma}$ get cyclically 
interchanged when $\ta$ changes by $\pi/3$ due to the underlying triangular lattice structure. The periodicity of $C$ on the other hand is $2\pi/3$, and its sign changes
when $\theta$ changes by $\pi$. 
In Fig.~\ref{fig:figure5} (a) for $a_2 =30$, we can see that there are interesting points at $\ta = \pi/3$ and $2\pi/3$ where the three two-spin 
coupling parameters have the same value. The value of $C$ at these points is also the largest in magnitude, equal to about $0.15$. In 
Fig.~\ref{fig:figure5} (b), we see ranges of $\theta$ where the magnitude of $C$ is greater than one of the nearest-neighbor couplings. In Figs.~\ref{fig:figure5} (c) and \ref{fig:figure5} (d) we notice that one of the coupling parameters is much larger than the other two. We also observe in Figs.~\ref{fig:figure5} (c) and \ref{fig:figure5} (d) that one or two of the coupling parameters has almost the same value over a range of $\theta$.

We note in Fig.~\ref{fig:figure5} that whenever $\theta$ is equal to 
$(2 \pi n/3) \pm \pi/6$ (where $n$ is an integer), two of the $J$'s are 
equal and $C = 0$. This is because for these values of $\theta$, the 
electric field is perpendicular to one of the sides of each triangle.
Then the system is invariant under a reflection about the direction 
of the electric field. Hence the two $J$'s which are related by the
reflection must be equal, and $C$ must vanish since the chiral 
three-spin term is odd under the reflection.

\begin{figure}[htb]
\centering
\includegraphics[width=0.5\textwidth]{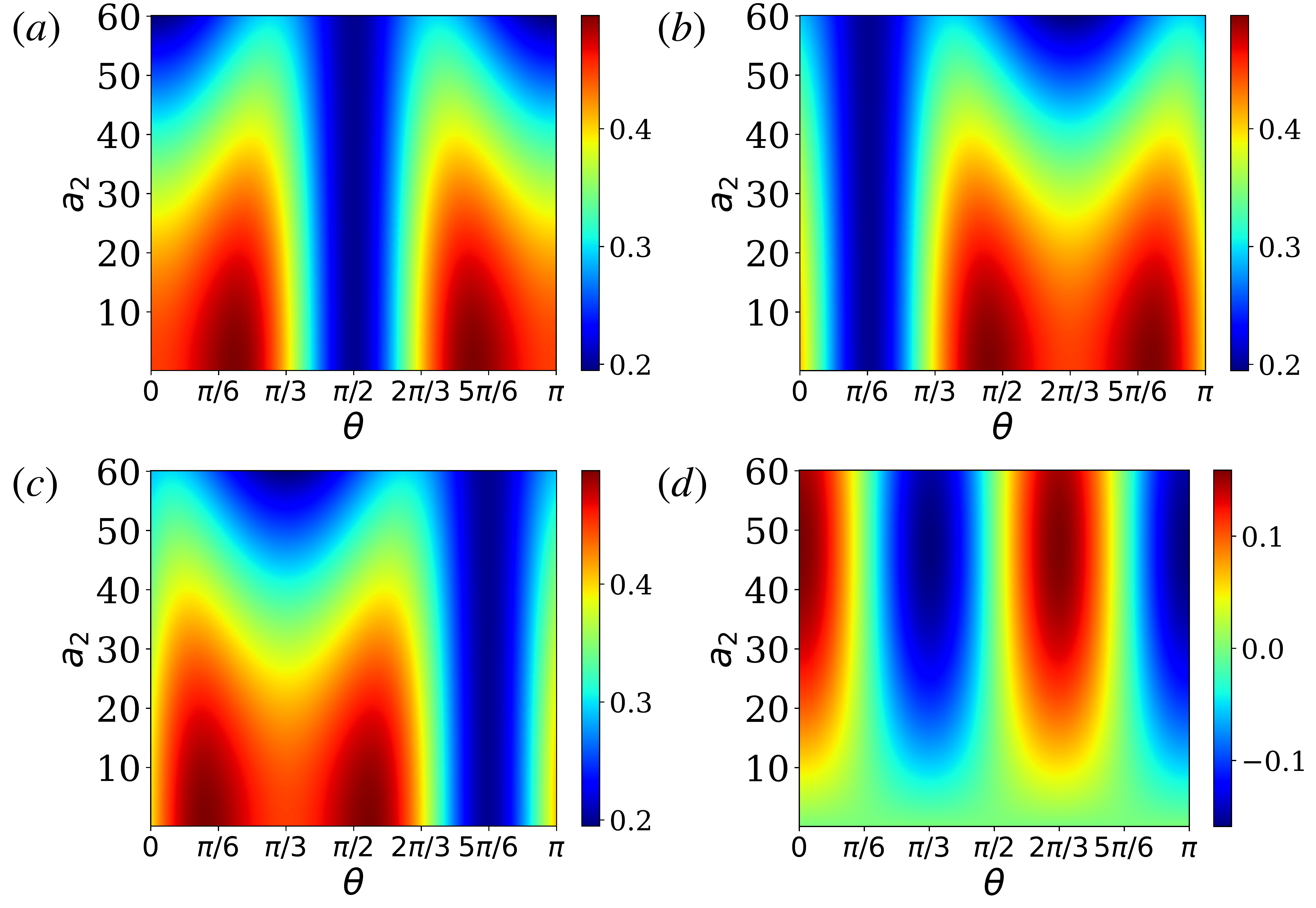}
\caption{Plots of $J_{\al}$, $J_{\beta}$, $J_{\gamma}$ and $C$ as functions of $\theta$ and $a_2$, for $g=1$, $U=20$, $\omega=17$, and $a_1$= 35. 
The plots show the behaviors of (a) $J_{\al}$, (b) $J_{\beta}$, (c) $J_{\gamma}$, and (d) $C$.} 
\label{fig:figure6}
\end{figure}

The variation of the four couplings as a function of both $a_2$ and $\theta$ is 
shown in Fig.~\ref{fig:figure6}, where we have fixed $g=1$, $U=20$, $\omega=17$, and 
$a_1 = 35$. We have varied $a_2$ from $0$ to $60$ and $\theta$ from $0$ 
to $\pi$. This interval of $\theta$ is sufficient to clearly show the periodicity 
of $J_{\al}$, $J_{\beta}$, $J_{\gamma}$ and $C$. We again see from these plots 
that $J_{\al}$, $J_{\beta}$, and $J_{\gamma}$ get cyclically interchanged as
$\theta$ changes by $\pi/3$, while $C$ has a period of $2\pi/3$.

\begin{figure}[htb]
\centering
\includegraphics[width=0.45\textwidth]{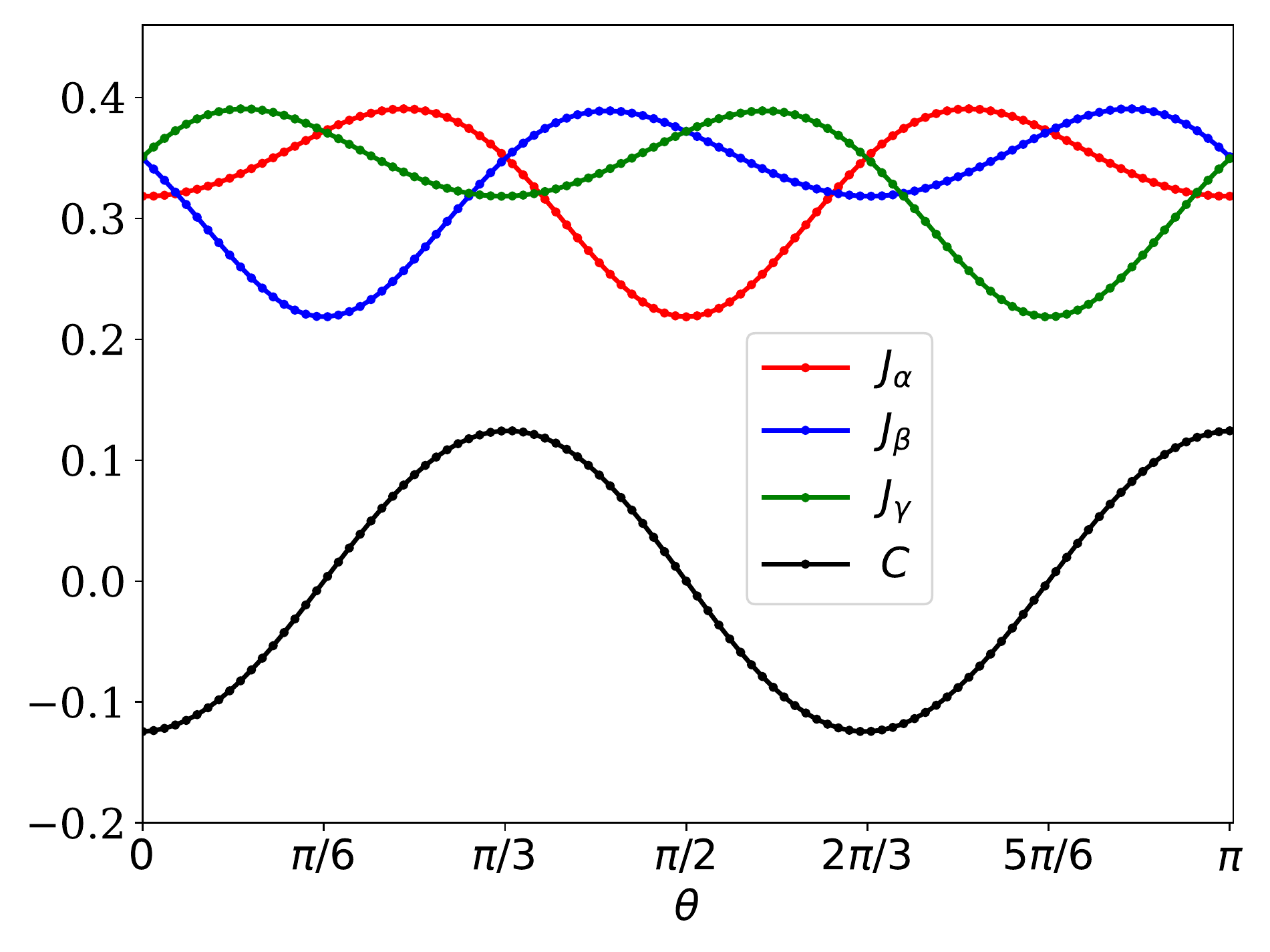}
\caption{Plots of $J_{\al}$, $J_{\beta}$, $J_{\gamma}$ and $C$ as functions of $\theta$, for $g=1$, $U=20$, $\omega=17$, $a_1$= 35 and $a_2 = 30$,
obtained by a numerical calculation of the Floquet operator for a single
triangle with two spin-up electrons and one spin-down electron, and truncating its
eigenvalues and eigenvectors to states with only singly occupied sites. The figure
shows a good match with Fig.~\ref{fig:figure5} (a) obtained from the third-order effective Hamiltonian presented in Eq.~\eqref{forms}.} 
\label{fig:figure7}
\end{figure}

In Fig.~\ref{fig:figure7}, we show plots of the couplings as 
functions of $\theta$ obtained directly from the Floquet operator for 
$g=1$, $U=20$, $\omega=17$, and $a_1 = 35$, and $a_2 = 30$. This 
calculation is done as follows. We consider a triangle of three sites
containing three electrons, two with spin-up and one with spin-down; 
the basis states are shown in Fig.~\ref{fig:figure2}. We first calculate 
the Floquet operator $U_T$ in Eq.~\ref{ut} by discretizing the 
time $t$ and multiplying $N$ terms with time steps $\Delta t = T/N$ 
(we have taken $N=201$). We then examine the nine eigenstates of $U_T$ and
find which three of them have the smallest amplitudes of the states with
doubly occupied sites (the last six states in Fig.~\ref{fig:figure2}). 
We truncate these eigenstates to states with only singly occupied sites
(the first three states in Fig.~\ref{fig:figure2}) and carry out a Gram-Schmidt orthogonalization to obtain three states $| \psi_i \rangle$, $i=1,2,3$
(the orthogonalization alters the states only slightly since the amplitudes 
of the six states which have been left out are small). Using the states 
$| \psi_i \rangle$ and their corresponding Floquet quasienergies $\ep_i$ 
(these lie near zero since we have chosen $g$ to be much smaller 
than $U$ and $\omega$), we construct the effective Hamiltonian 
\beq H^{eff}_{\bigtriangleup} ~=~ \sum_{i=1}^3 ~\ep_i ~|\psi_i \rangle \langle 
\psi_i |. \eeq 
We then fit this to the form given in Eq.~\eqref{Spinham} to extract $J_\al$, $J_\beta$, $J_\ga$ and $C$. These are plotted in Fig.~\ref{fig:figure7}. Note 
that these plots also exhibit the symmetries mentioned above for 
various values of $\theta$. A comparison between 
Figs.~\ref{fig:figure5} (a) and \ref{fig:figure7} 
shows a good match, indicating that the results that we have 
obtained from the third-order effective Hamiltonian in
Eq.~\eqref{forms} agree well with exact numerical calculations.

\subsection{Classification of different phases using static spin structure function}
\label{sec5b}

The numerical values of the parameters $J_{\al}$, $J_{\beta}$, $J_{\gamma}$ and $C$ obtained for different driving parameters give us an idea of the ranges of values that they can have. To find the different phases of the system using ED, we have varied the parameters $J_{\beta}$, $J_{\gamma}$ and $C$ independently, yet consistent with the values obtained by
driving. (We have fixed $J_{\al} = 1$ for convenience). 
For each set of parameters, we have calculated the static two-spin correlation function in the ground state, given by the formula
$C(\vec{r}_{n},\vec{r}_{0}) = C(\vec{r}_{n}-\vec{r}_{0}) = \braket{\vec{S}_{0}~\cdot~ \vec{S}_{n}} $. This correlation function 
tells us the kind of order present in the ground state. In the
case of ordered ground states, the SSSF, defined as the Fourier transform of the correlation function 
\beq S(\vec{q}) ~=~ \frac{1}{\sqrt{N}} ~\sum_{\vec{r}_{n}} e^{-i \vec{q} \cdot (\vec{r}_{n}-\vec{r}_{0})}~ C(\vec{r}_{n}-\vec{r}_{0}) \eeq
(where $N$ is the number of lattice sites) peaks sharply at particular points $\vec{q}$ in the Brillouin zone. The positions of these peaks 
indicates the nature of the order. On the other hand, SSSF does not have 
a well-defined peak at any point in the Brillouin zone for a disordered ground state.

For our choice of the lattice vectors $\vec{u}$ and $\vec{v}$, the Brillouin zone in reciprocal space in spanned by the reciprocal lattice vectors $\vec{q_{x}}$ and $\vec{q_{y}}$ which run from $0$ to $2\pi$ and $0$ to $4\pi/\sqrt{3}$, 
respectively, as shown in Fig.~\ref{fig:figure8}.
From the SSSF calculations we have classified a total of seven possible phases, of which four are ordered and the other
three are disordered. We have shown representative plots of SSSF for each of these phases in Fig.~\ref{fig:figure8}. To summarize the
ordered phases, we have shown the $\vq$-points where the SSSF has peaks for the different ordered phases in the 
top left figure in Fig.~\ref{fig:figure8}.

In Sec.~\ref{sec5d}, we will confirm the different phases obtained from 
the SSSF  using other quantities like the fidelity susceptibility and 
real-space correlation values at large distances.

\vspace*{.4cm}

\subsection{Ground state phase diagram}
\label{sec5c}

We now present the phase diagram as a function of $J_\beta/J_{\al}$ and $J_\gamma /J_{\al}$ in Fig.~\ref{fig:figure9}. We saw in
Sec.~\ref{sec5a} that the values obtained for $C$ by 
driving is usually small compared to $(J_\al, J_\beta$ and $J_\ga$, except in some 
small regions where the $C$ is comparable to one of the couplings. Further,
we have found numerically that the SSSF calculated for the ground state 
does not change significantly on including the values of $C$ 
obtained by driving even when it is comparable to one of the 
two-spin couplings. Hence the phase diagram is practically 
independent of the value of $C$. We have therefore set $C=0$ in 
Fig.~\ref{fig:figure9}.

\subsection{Verification of different phases: fidelity susceptibility, minimum real-space correlation function and energy levels}
\label{sec5d}

Quantum phase transitions can often be captured by looking at the ground-state fidelity as a function of the parameters of the system. Fidelity is a concept borrowed from quantum information theory. It is defined as
\beq \mathcal{F}(\lambda) ~=~ |\braket{\psi_{0}(\lambda) | \psi_{0}(\lambda+ \delta \lambda)}|, \eeq
where $\ket{\psi_{0}(\lambda)}$ and  $\ket{\psi_{0}(\lambda+ \delta \lambda)}$ are the ground states of the many-body Hamiltonian $H$ with slightly different parameters $\lambda$ and $ \lambda+ \delta \lambda$ respectively. For a fixed and small value of $\delta \lambda$,
$\mathcal{F}$ generally exhibits a prominent dip whenever a phase
transition occurs between $\lambda$ and $ \lambda+ \delta \lambda$. 
As a result, the second derivative of the fidelity with respect
to $\lambda$ usually shows large changes near a quantum critical point.
This leads us to define another measure 
called the fidelity susceptibility~\cite{you,wang}
\begin{equation}
\chi_{F}(\lambda) ~=~ - \left. \frac{\partial^{2} \ln 
\mathcal{F}}{\partial (\de \lambda)^2} \right|_{\delta 
\lambda \rightarrow 0}. \end{equation}

\begin{widetext}
\begin{figure*}[htb]
\includegraphics[width=\textwidth]{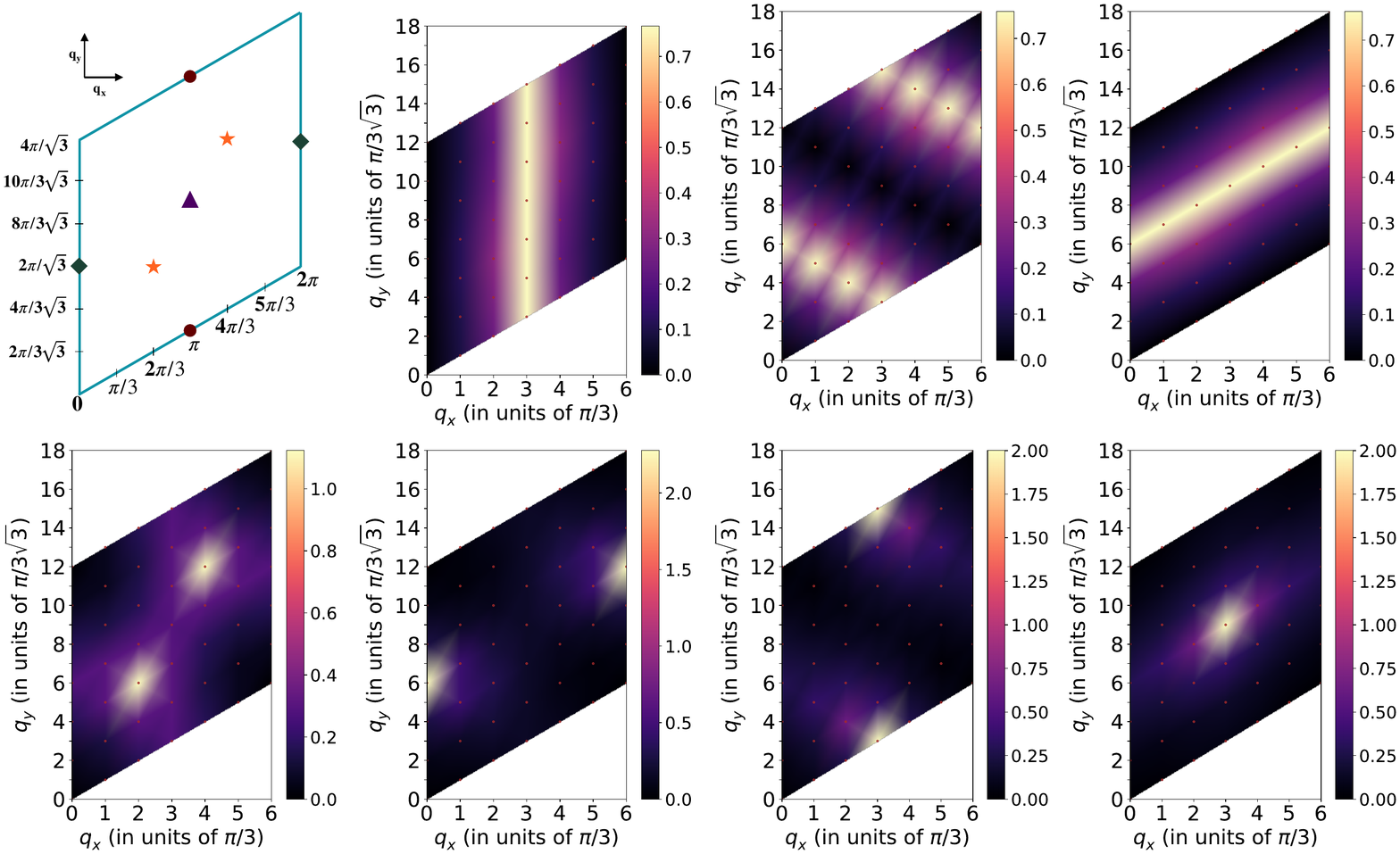}
\caption{The first figure in the top row shows the locations of the 
peaks in the SSSF for different ordered phases in the Brillouin
zone of the triangular lattice. The peaks at 
$(\pi/3,2\pi/\sqrt{3})$ and $(4\pi/3, 4\pi/\sqrt{3})$ indicated by 
\includegraphics[width=0.02\textwidth]{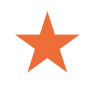} correspond to the peaks in SSSF in the spiral phase, as shown in the first 
figure in the bottom row. The peaks located at the points marked at
$(0,2\pi/{\sqrt 3})$ and $(2\pi,4\pi/{\sqrt 3})$ shown by 
\includegraphics[width=0.02\textwidth]{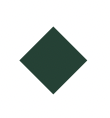}, 
$(\pi,\pi/{\sqrt 3})$ and $(\pi,5\pi/{\sqrt 3})$ shown by 
\includegraphics[width=0.02\textwidth]{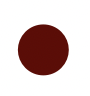}, 
and $(\pi,3\pi/{\sqrt 3})$ represented by
\includegraphics[width=0.02\textwidth]{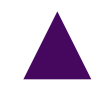} correspond 
to the collinear phases called 
stripe-$\vec{u}$, stripe-$\vec{v}$ and 
stripe-$\vec{w}$, and the last three figures in the bottom row show 
plots of the SSSF in these three phases respectively. The SSSFs for 
the spin-liquid phases are shown in the last three figures in the 
top row. 
We see that the largest values of the SSSF are spread out over many 
$\vec{q}$ points in the spin-liquid phases. In the third figure in 
the top row, the apparent fringes are only due to the interpolation 
scheme used while plotting, and the SSSF values are actually 
highest all along the two parallel lines in this figure.} 
\label{fig:figure8} 
\end{figure*} 
\end{widetext}

\begin{widetext}
\begin{figure*}[htb]
\includegraphics[width=\textwidth]{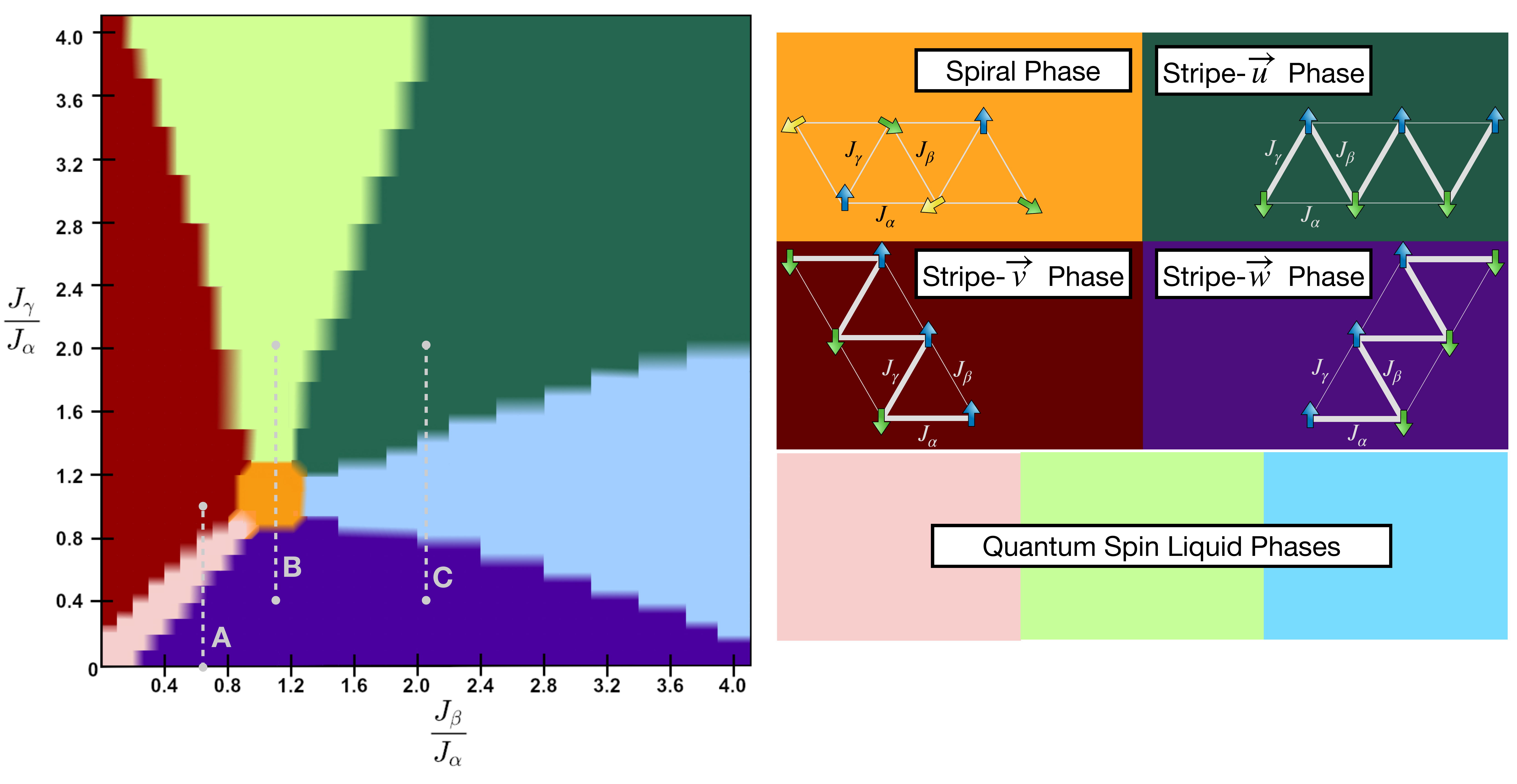}
\caption{The phase diagram of the effective spin model as a function of $J_{\beta}/J_{\al}$ and $J_{\gamma}/J_{\al}$, for $C=0$. The 
different phases classified by the SSSF are shown in the figure on the right. We have seven phases altogether, of which four are ordered and 
three are disordered or spin-liquid phases. The stripe phases are 
separated from each other by intermediate spin-liquid phases as shown
in the figure on the left. The phases boundaries remain the same on interchanging $J_{\beta}$ and $J_{\gamma}$, only the directions of the 
spin configurations change as expected from the geometry of the 
triangular lattice.  We have marked three lines $A$, $B$ and $C$ in 
the phase diagram which 
encompass all the seven phases. We have calculated the fidelity 
susceptibility, real-space correlation function, and energy gaps 
along these lines to verify the phase boundaries obtained using
SSSF. Note that we used a 
resolution of $0.1$ in both directions while constructing the phase 
diagram; this explains the discrete stair-like structure of the phase boundaries.} 
\label{fig:figure9}
\end{figure*}
\end{widetext}

\begin{widetext}
\begin{figure*}[htb]
\includegraphics[width=\textwidth]{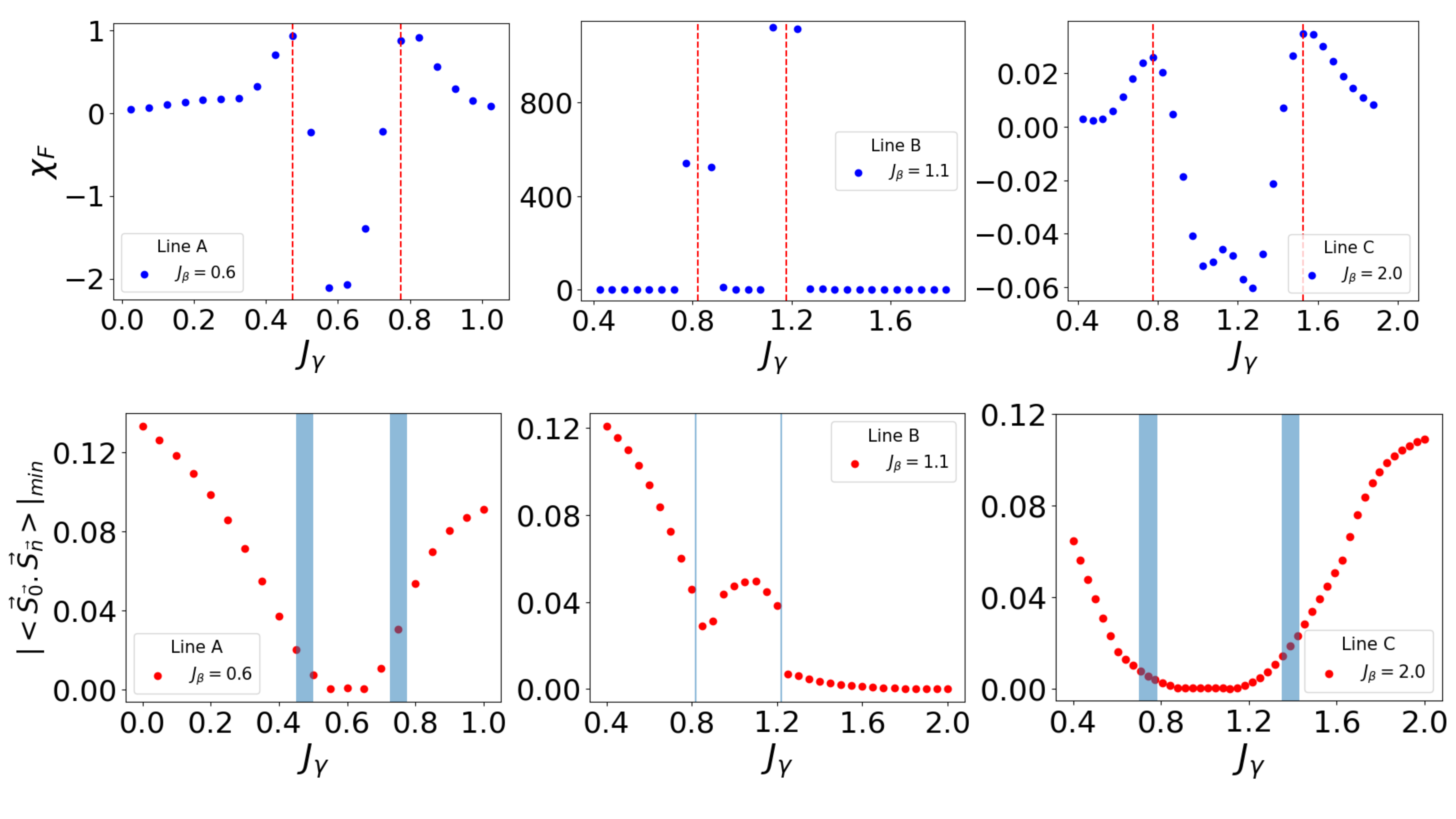}
\caption{Plots of the fidelity susceptibility (top row) and the minimum
value of the real-space correlation function (bottom row) along the
lines $A$, $B$ and $C$ shown in Fig.~\ref{fig:figure9}. The fidelity 
susceptibility shows significant changes at all the phase transition 
lines. However, the change is
extremely large at the transitions on line $B$; note that the scale
of the $y$-axis in the plot for line $B$ is quite different from the 
scales for lines $A$ and $C$. Similarly, the minimum value of the
real-space correlation function seems to change continuously at the 
transitions on lines $A$ and $C$ but discontinuously at the 
transitions on line $B$.} 
\label{fig:figure10} 
\end{figure*}
\end{widetext}

At a critical point $\chi_{F}$ generally shows a maximum or a 
divergence. To confirm the 
phase boundaries shown in Fig.~\ref{fig:figure9}, we have chosen three vertical lines $A$, $B$ and $C$ which, taken together, cover all the seven phases, and we have
calculated $\chi_{F}$ along these lines. The top row in Fig.~\ref{fig:figure10} shows plots of $\chi_{F}$ as a function of the parameter $J_{\gamma}$. The locations of phase boundaries obtained 
from the peaks in $\chi_{F}$ and from the SSSF calculations agree well 
with each other. For line $A$, with $J_{\beta}= 0.6$, we see from Fig.~\ref{fig:figure9} that it passes from stripe-$\vec{w}$ through a spin liquid to 
stripe-$\vec{v}$. The fidelity susceptibility along this line shows two maxima at $J_{\gamma}= 0.475$ and $0.77$ which mark the phase 
transitions to and from the spin-liquid 
phase. Similarly line $C$ shows maxima in $\chi_F$ at 
$J_{\gamma}=0.775$ and $1.5$ indicating a similar phase transition 
from stripe-$\vec{w}$ through a spin-liquid to stripe-$\vec{u}$. 
On line $B$, however, we find a divergence 
in $\chi_F$ at $J_{\gamma}=0.82$ and $1.18$. These points match the
phase transitions seen in Fig.~\ref{fig:figure9} when the system 
goes from stripe-$\vec{w}$ to the spiral phase to a spin liquid. 
The divergences in $\chi_F$ in this case suggests discontinuous 
transitions and these may occur because the transitions here are from 
a spiral phase (which has a very
different kind of structure as shown by the SSSF in 
Fig.~\ref{fig:figure8}) to a striped phase or to a spin liquid.

We have also used the minimum value 
of the two-spin correlation function (the value at the largest possible
distance between two spins, namely, half-way across the system) 
as a tool to distinguish between different phases. In the ordered 
phases, $\braket{\vec{S}_{\vec 0}~\cdot~ \vec{S}_{\vn}}$ at large 
separation $|\vn|$ goes to a finite value , while in a spin-liquid 
phase the correlation approaches zero quickly with increasing 
separation. The minimum value of the two-spin correlation function 
captures the correlation at the largest distance possible in our $6 
\times 6$ lattice.  In the bottom row of 
Fig.~\ref{fig:figure10} we have shown the variation of the minimum 
correlation versus $J_\gamma$ along the lines $A$, $B$ 
and $C$. In each of the lines we see a rapid 
drop in the value whenever it is in a spin-liquid phase. The minimum 
correlation is of the order of
$10^{-1}$ -- $10^{-2}$ in the ordered phases and of the order 
of $10^{-3}$ -- $10^{-4}$ in the spin-liquid phases. The phase 
boundaries obtained by this method are not as sharp as ones obtained 
from $\chi_F$; however, they still agree quite well with each other.
Both the fidelity susceptibility and the minimum value of the 
real-space correlation function in Figs.~\ref{fig:figure10} suggest 
that transitions between a striped phase and a spin liquid are 
continuous (lines $A$ and $C$), while a transition between the 
spiral phase and either a striped phase or a spin liquid is 
discontinuous (line $B$).

\begin{widetext}
\begin{figure*}[htb]
\includegraphics[width=\textwidth]{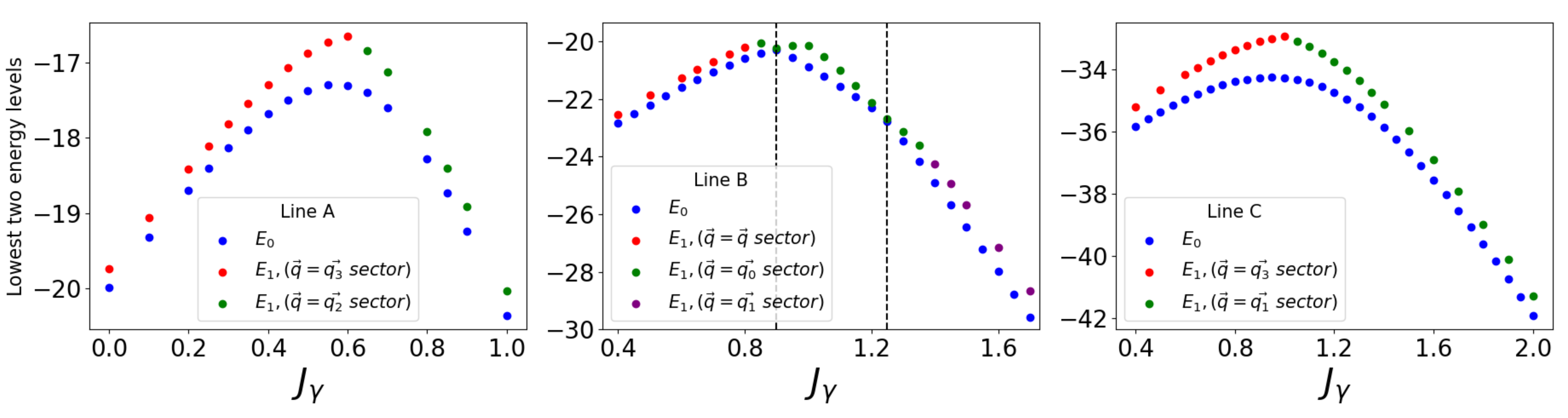}
\caption{Plots of the ground state and first excited state energies along
the lines $A$, $B$ and $C$ shown in Fig.~\ref{fig:figure9}. The gap 
between the ground state and first excited state
remains finite all along lines $A$ and $C$ but becomes close to zero at the transitions on line $B$. The ground state always lies in the symmetry sector with even parity ($P=1$), even spin-inversion ($Z=1$) and momentum $(0,0)$. However, the symmetry sectors of the excited state changes as $J_{\gamma}$ is varied. For line A, the excited state has $P=-1$ and $Z=-1$ with the momentum sector changing from $\vec{q_3}= (\pi, 3\pi/\sqrt{3})$ to $\vec{q_2}= (\pi, \pi/\sqrt{3})$ at $J_{\gamma}= 0.6$. Similarly for line C at $J_{\gamma}=1.0$, the momentum sector switches from $\vec{q_3}$ to $\vec{q_1}= (0, 2\pi /\sqrt{3})$ with $P=-1$ and $Z=-1$. Along line B, the excited state changes 
once from $\vec{q} = 
\vec{q}_{3}$, $P=1$, $Z=-1$ to $\vec{q} = \vec{q}_{0} = (0,0)$, $P=1$, 
$Z=1$, and then again to $\vec{q} = \vec{q}_{1}$, $P=-1$, $Z=-1$. These changes occur in the vicinity of the energy gap closing points.}
\label{fig:figure11}
\end{figure*}
\end{widetext}

\begin{figure}[H]
\centering
\includegraphics[width=0.5\textwidth]{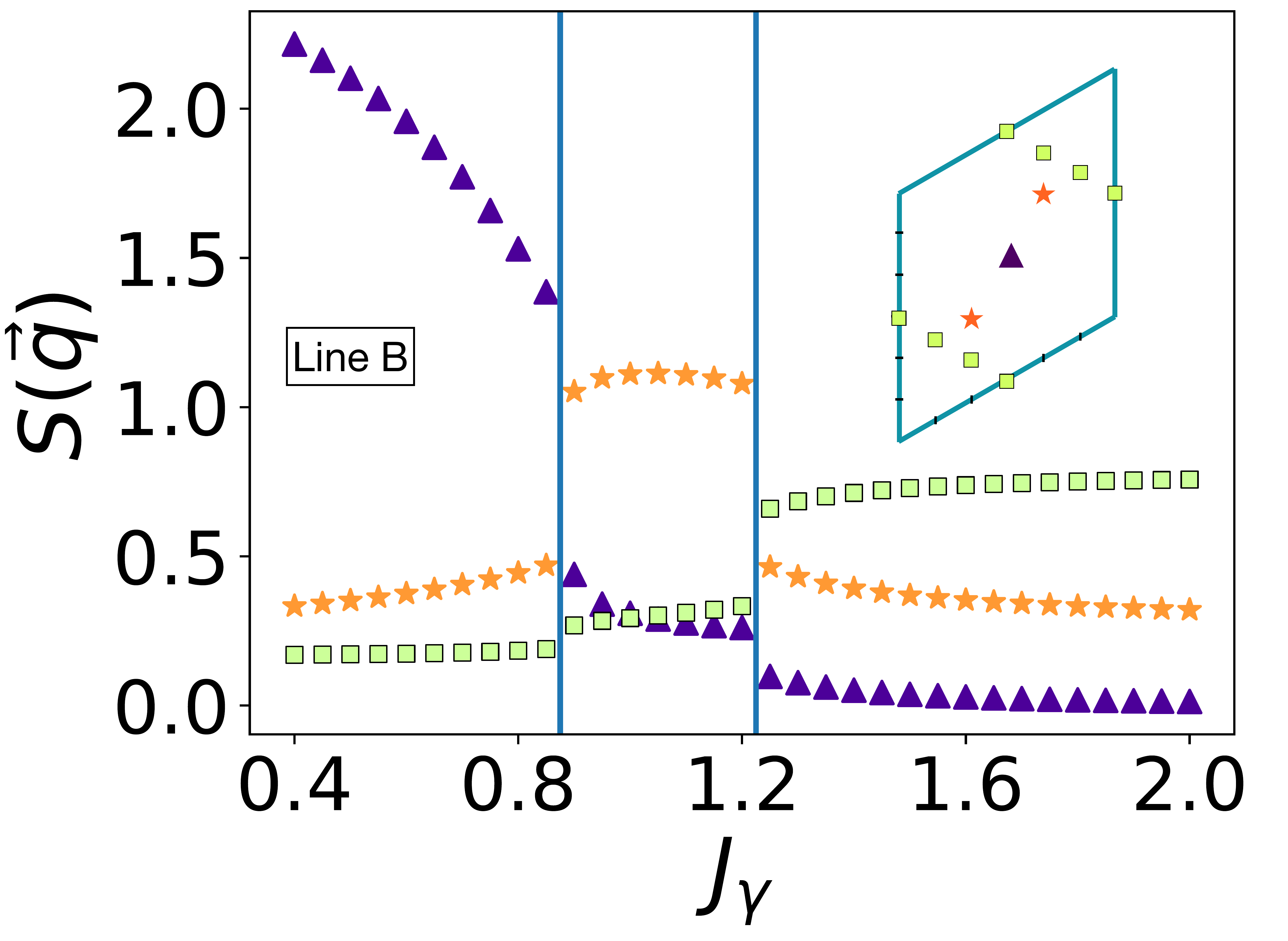}
\caption{SSSF plotted at different values of $\vq$ in the Brillouin zone 
which are shown in the inset.
The SSSF is plotted along line $B$ in Fig.~\ref{fig:figure9}, 
with $J_{\alpha}=1$ and $J_{\beta}=1.1$ held fixed. Line B goes across three 
phases as $J_{\gamma}$ is increased, and the SSSF shows 
corresponding transitions in its values at the two 
phase boundaries. $S(\vec{q})$ has the maximum value for the stripe-$\vec{v}$ 
phase at $\vq =$ \includegraphics[width=0.015\textwidth]{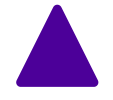} from 
$J_{\gamma}= 0$ to $J_{\gamma} 
\simeq 0.82$, for the spiral phase at $\vq =$ 
\includegraphics[width=0.015\textwidth]{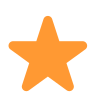} from $J_\ga 
\simeq 0.82$ to $J_{\gamma} \simeq 1.2$, and for the spin-liquid phase at 
$\vq =$ \includegraphics[width=0.015\textwidth]{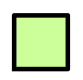} from $J_{\gamma} 
\simeq 1.2$ onwards.} 
\label{fig:figure12}
\end{figure}

We have also studied the first excited state energy along the lines 
$A$, $B$ and $C$ as shown in Fig.~\ref{fig:figure11}. While the 
ground state has momentum $(0,0)$ and is in the even parity ($P=1$) 
and 
even spin-inversion sector ($Z=1$) for all values of the parameters, 
the excited state has different momenta and lies in different parity 
sectors in different regions in the parameter space. Along line $A$, 
which is at $J_\beta = 0.6$, we find that as we vary $J_\gamma$ from 
$0$ to $1$, the excited state is in the momentum sector $\vec{q} = 
\vec{q}_{3} = (\pi,3\pi/\sqrt{3})$ with $P=1$ and $Z=-1$. We then have 
a transition at $J_\gamma = J_\beta = 0.6$ after which the excited 
state has momentum $\vec{q} = \vec{q}_{2} = (\pi,\pi/\sqrt{3})$ with 
$P=-1$ and $Z=-1$. The plot for line $C$ is quite similar. Here we have fixed $J_\beta = 2.0$ and $J_\gamma$ is varied from $0.4$ to $2$. The 
symmetry sector of the excited state changes from $\vec{q} = \vec{q}_{3} = (\pi,3\pi/\sqrt{3})$, $P=1$, $Z=-1 $ to  $\vec{q} = \vec{q}_{1} = (0,2\pi/\sqrt{3})$, $P=-1$, $Z=-1$ at $J_\gamma = 1 $. 
Along both lines $A$ and $C$, we cannot comment on the nature of 
the transition 
from ordered to spin-liquid or vice-versa by looking at the energy gap 
between the ground state and first excited state. However, we expect 
that in the stripe phases the energy gap will vanish in the 
thermodynamic limit since
spin-wave theory about the classical stripe phases predicts a gapless dispersion. For line $B$, we have fixed $J_\beta = 1.1$ and varied $J_\gamma$ from $0.4$ to $2$. This line contains the spiral phase and the energy gap closes at its phase boundaries. Along this line the 
excited state changes its symmetry sector twice, once from $\vec{q} = 
\vec{q}_{3}$, $P=1$, $Z=-1$ to $\vec{q} = \vec{q}_{0} = (0,0)$, $P=1$, 
$Z=1$, and then again to $\vec{q} = \vec{q}_{1}$, $P=-1$, $Z=-1$. These changes occur near the transition from stripe-$w$ phase to the spiral phase and from the spiral phase to one of the spin-liquid phases respectively. For this line also we expect that the energy gap will vanish in the thermodynamic limit for the ordered phases. Moreover, the spin-liquid phase on line $B$ has a small gap
(of the order of or smaller than the inverse system size); hence it
is like to be gapless in the thermodynamic limit. For line B, we have shown in Fig.~\ref{fig:figure12} how $S(\vec{q})$ varies for the $\vq$-points in the three phases that the line covers. We see that the maximum value of $S_{\vec{q}}$ is at $\vec{q}= (\pi, 3\pi/\sqrt{3})$ corresponding to the stripe-$\vec{w}$ phase until $J_{\gamma}= 0.82$, and at $(2\pi/3, 2\pi/\sqrt{3})$ and $(4\pi/3,4\pi/\sqrt{3})$ corresponding to the spiral phase until $J_{\gamma}=1.2$. Beyond this we find that the maximum value of $S_{\vec{q}}$ is spread across the two lines as shown in the inset of Fig.~\ref{fig:figure12} which correspond to a spin-liquid phase. This further confirms the phase boundaries for this line.

We note that all the three spin-liquid phases are similar in the sense that one can 
obtain one from the other by permuting or exchanging the parameters $J_\al, ~J_\beta$ and $J_\ga$. Hence, since the spin-liquid phase on
line B appears to be gapless (Fig.~\ref{fig:figure11}), all the spin-liquid
phases are likely to be gapless.

\subsection{Effect of $C$}
\label{sec5e}

\begin{figure}[htb]
\centering
\includegraphics[width=0.5\textwidth]{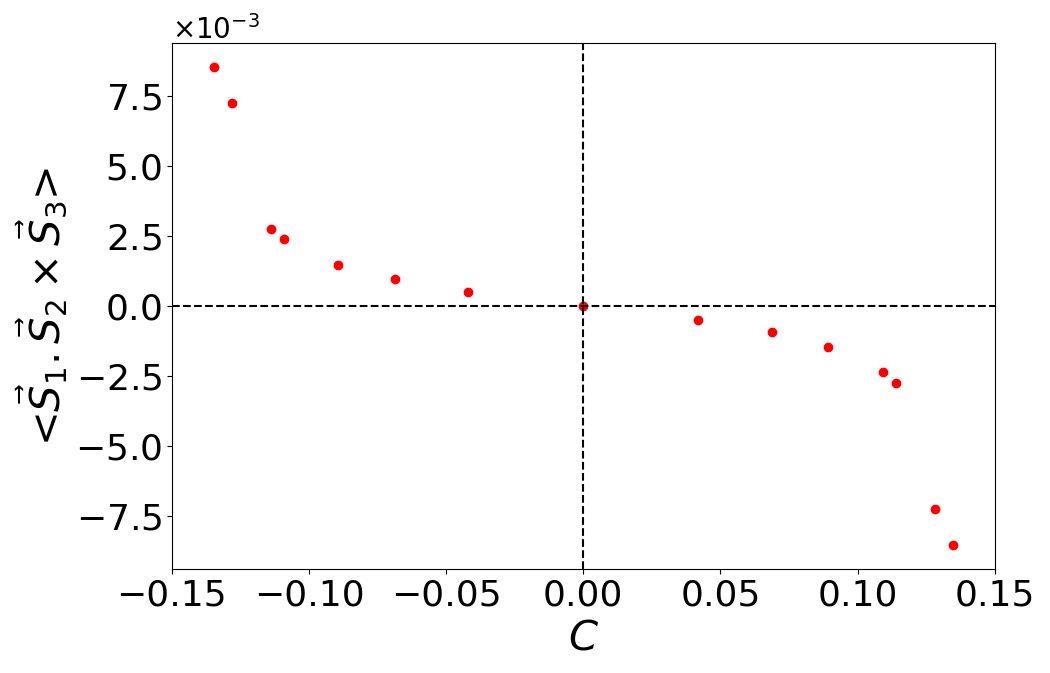}
\caption{Plot of chiral three-spin term $\braket{\vec{S}_1 \cdot 
\vec{S}_2 \times \vec{S}_3}$ for any triangle versus $C$, for a system
with $g=1$, $U=20$, $\om =17$, $a_1 = 35$ and $a_2 = 30$.} 
\label{fig:figure13}
\end{figure}

Although the chiral three-spin term with coefficient $C$ does not seem to
play an important role in the phase transitions between ordered and 
spin-liquid phases, it does have some effect on the ground state. The classical calculation in the large-$S$ limit suggests that the effect of the three-spin term is to make the spin configuration non-coplanar
in every triangle. In addition, the staggered structure of the
three-spin term for up- and down-pointing triangles ensures that the 
energy can be minimized for all triangles simultaneously by having a 
particular non-coplanar three-sublattice order. We can measure this
non-coplanarity using an order parameter given by
the ground state expectation value of the chiral
three-spin term $\vec{S}_1 \cdot \vec{S}_2 \times \vec{S}_3$ on 
any triangle in the lattice, taken in an anticlockwise (clockwise) 
sense for up-pointing (down-pointing) triangles. This is shown in Fig.~\ref{fig:figure13} for values of $C$ (and the corresponding values
of $J_\al, ~J_\beta$ and $J_\ga$) obtained as a function of $\ta$ in
Fig.~\ref{fig:figure5} (a). As expected, we find that this order parameter
is an odd function of $C$ and it vanishes if $C=0$.

\section{Discussion}
\label{sec6}

In this paper we have studied the effects of a periodically varying 
in-plane electric field on the Hubbard model at half-filling on a 
triangular lattice. In the limit that the nearest-neighbor 
hopping amplitude $g$ is much smaller 
than the interaction strength $U$ and the driving frequency $\om$, we have 
used a Floquet perturbation theory to derive the effective Hamiltonian up to
order $g^3$ in the spin sector, namely, the sector of states in which all sites 
are singly occupied. Assuming that there is no resonance, i.e., $U$ is not 
close to an integer multiple of $\om$, we find that the
Hamiltonian is given by a sum of nearest-neighbor Heisenberg interactions 
at orders $g^2$ and $g^3$, and, if the electric field is not time-reversal 
symmetric, a chiral three-spin interaction on each triangle at order $g^3$.
Indeed, the reason we chose to study the Hubbard model on a triangular lattice 
is that it is known that a magnetic field which is perpendicular to the plane 
of the lattice gives rise, at order $g^3$ in time-independent perturbation 
theory, to
a chiral three-spin interaction with a coefficient which depends on the 
magnetic flux passing through each triangle~\cite{chitra}. Thus an 
oscillating electric field in our model can simulate the effect of a 
magnetic flux in a time-independent system. Interestingly however, while the
sign of the three-spin term written in the anticlockwise direction
is the same for up- and down-pointing triangles in the time-independent
magnetic flux problem, the sign is opposite in the two kinds of triangles
in our periodically driven problem. 

In our numerical calculations we have chosen the oscillating electric
field to be linearly polarized with two different frequencies in order to
break time-reversal symmetry. This is in contrast to earlier work which 
showed that chiral three-spin terms can be generated when circularly polarized 
radiation with a single frequency is applied to certain frustrated Mott 
insulators, and these terms appear at fourth order 
in $g$~\cite{claassen,kitamura,bostrom,torre}. It has also been shown 
that partially polarized and unpolarized radiation with a single
frequency can generate chiral three-spin and other multi-spin terms at 
fourth order in $g$ in various Mott insulators~\cite{quito1,quito2}.

The coefficients of the two-spin Heisenberg interactions in our effective 
spin Hamiltonian are found to have different values, $J_\al, ~J_\beta, ~J_\ga$,
for bonds pointing along the three different directions on the triangular 
lattice. The values of $J_\al, ~J_\beta, ~J_\ga$ and the coefficient $C$
of the three-spin term depend on all the driving parameters such as the
amplitude and frequency of driving and the direction of the electric field.
(Typically, $C$ is found to be smaller than $J_\al, ~J_\beta$ and $J_\ga$).
We thus obtain an interesting spin model whose parameters can all be tuned 
by the driving. We then study this spin model in detail. We first carry
out a classical analysis (by taking the spin at site to be very large
instead of 1/2) to find the possible ground state spin configurations. 
Depending on the spin model parameters we find that there 
are three collinear and one
coplanar ordered state. We then use spin-wave theory to find the excitation
spectrum about one of the collinear ground states; we find that this 
theory breaks down close to the transition to a different phase, which
hints at the possibility of some disordered phases. 

Next, we use ED to numerically study systems with
36 sites with periodic boundary conditions. We concentrate on the ground state
and use various symmetries of the system to reduce the Hilbert space dimension,
by working in the sector with zero momentum in both directions, zero
total spin in the $z$ direction, even spin inversion, and, if $C=0$,
even spatial inversion. After finding the ground state, we look at the
two-spin correlation function in real space, the SSSF, and the
fidelity susceptibility. We also look at the energies of the ground state
and the lowest excited state. Putting together all this information, we
find a rich ground state phase diagram consisting of three collinear 
and one coplanar ordered phase (in agreement with the classical analysis)
as well as three disordered phases. We find transitions between the
coplanar phase and all the other six phases. In each of the ordered phases, 
the SSSF in momentum space has a peak at one or two points in the Brillouin
zone, while in each of the disordered phases, the SSSF is large along some 
lines. Away from the phase transition lines, the peak values of the SSSF 
are significantly smaller in the disordered phases compared to the ordered 
phases. In real space, the two-spin correlation function at the largest 
possible separation (namely, between two points separated by half the
system size in both directions) is found to be finite in the ordered
phases and very small in the disordered phase; this is expected for systems 
with and without long-range order respectively. The ground state fidelity
susceptibility is found shows significant changes whenever a 
phase transition
line is crossed; the changes are much larger at the transitions between the
coplanar phase and the other phases, compared to the changes which occur
at transitions between collinear ordered and disordered phases. This is
related to the observation that the ground state and first excited state
remain well separated in energy for at transitions between collinear 
ordered and disordered phases but come very close to each other at transitions
between the coplanar phase and the other phases. For $C=0$, our phase diagram 
is in broad
agreement with the ones reported in Refs.~\onlinecite{weng,yunoki,hauke}.
Finally, we have found that the values of $C$ that are typically generated by 
periodic driving are not large enough to significantly modify the 
ground state phase diagram.

The effective Hamiltonian that we have studied in this paper applies
only to the spin sector where each site is occupied by a single electron. 
This description is valid in a prethermal regime, and it is known 
that in systems with short-range interactions, the duration of this regime is 
exponentially large when the frequency is much larger than the 
hopping~\cite{abanin,mori,tran}, as we have
assumed in our numerical calculations.
Eventually, after an exponentially long time, the periodic driving is 
expected to heat up our system to infinite temperature where all states 
are equally probable; then the analysis in this paper will break down.

In summary, our work proposes a way of simulating a tunable spin model by 
periodically driving a fermionic system with strong interactions. Earlier theoretical
works have studied the effects of chiral three-spin terms generated by
circularly polarized radiation applied to kagome Mott insulators such as
herbertsmithite~\cite{claassen,kitamura} and magnetic systems like 
CrI$_3$~\cite{bostrom}. The values of the Hubbard interaction $U$ and the 
photon energy $\hbar \om$ considered in these systems are typically of the 
order of 1 eV, and the ratio $U/g$ is about 20-30.

In this work we have considered a closed system which is not coupled to a 
thermal bath at some temperature. Coupling to a bath is generally expected to 
lead to a Floquet-Gibbs distribution of the states when a periodically driven 
system is not integrable~\cite{shirai1,shirai2}. The effects of a bath on our
system may be an interesting problem for future studies.

We would like to end by mentioning some of the recent experiments where spin-liquid 
and magnetically ordered phases have been realized on a triangular lattice. When 
ultracold bosonic atoms on a triangular optical lattice are periodically shaken in 
an elliptical manner~\cite{struck,eckardt}, it is found that the system is 
effectively governed by a spin model whose couplings can be tuned at will. This 
allows for the realization of various ordered and disordered phases at high enough 
temperatures. On the other hand, there are several magnetic materials like the 
organic salts Me$_{4-n}$Et$_{n}$P$n$[Pd(dmit)$_{2}$]$_{2}$,~\cite{scriven} 
TMTTF,~\cite{yoshimi} and BaAg$_{2}$Cu[VO$_{4}$]$_2$~\cite{tsirlin} where 
first-principle calculations have shown that they can be described by a 
triangular lattice antiferromagnet with spatially anisotropic exchange 
couplings similar to the ones studied in our paper.

\vspace{0.8cm}

\centerline{\bf Acknowledgments}
\vspace{0.5cm}

S.S. thanks MHRD, India for financial support through a PMRF.
D.S. acknowledges funding from SERB, India (JBR/2020/000043). We acknowledge Marin 
Bukov and Phillip Weinberg for helping us out in the use of their ED package 
QuSpin~\cite{quspin}
which was essential for this work. We thank Manish Jain and Prasad Hegde for the use 
of their clusters where the numerical calculations were carried out. We also thank 
S. Ramasesha, Krishnendu Sengupta, Subhro Bhattacharjee, Bhaskar Mukherjee, 
Shinjan Mandal, Dayasindhu Dey, and Niall Moran for useful discussions.

\end{document}